\documentclass[prx,twocolumn,showpacs,amsfonts,amsmath,floatfix,superscriptaddress]{revtex4-2}
\usepackage{graphicx}
\usepackage{blindtext}
\usepackage{eucal}
\usepackage{import}
\usepackage{color,verbatim,bbold,float,wrapfig}
\usepackage{amssymb,amsbsy,amsmath,mathrsfs}
\usepackage{caption}
\usepackage{times}
\usepackage{color}
\usepackage{setspace}
\usepackage{bm}% bold math.
\usepackage{braket}
\usepackage{commath}
\usepackage[dvipsnames]{xcolor}
\usepackage{amsmath}
\usepackage{amssymb}
\usepackage{csquotes}
\usepackage{enumitem}
\usepackage{float}
\usepackage{subfigure}
\usepackage{dsfont}
\usepackage{xcolor}
\usepackage{comment}
\usepackage[T1]{fontenc}
\usepackage{mathtools}
\usepackage{calc}
\usepackage{url}
\usepackage[colorlinks,allcolors=blue]{hyperref}
\usepackage{physics}
\usepackage{tikz}
\usetikzlibrary{shapes.geometric, arrows}

\tikzstyle{startstop} = [rectangle, rounded corners, minimum width=3cm, minimum height=0.5cm,text centered, draw=black, fill=red!30]
\tikzstyle{io} = [trapezium, trapezium left angle=70, trapezium right angle=110, minimum width=3cm, minimum height=1cm, text centered, text width = 6cm, draw=black, fill=blue!30,aspect = 0.2,trapezium stretches=true]
\tikzstyle{process} = [rectangle, minimum width=3cm, minimum height=1cm, text centered, draw=black, fill=orange!30]
\tikzstyle{process2} = [rectangle, minimum width=1cm, minimum height=1cm, text centered, draw=black, fill=orange!30]
\tikzstyle{decision} = [diamond, minimum width=3cm, minimum height=1cm, text centered, draw=black, fill=green!30,aspect=5]

\tikzstyle{arrow} = [thick,->,>=stealth]

\def\l{\left} 
\def\r{\right}
\def\nn{\nonumber}
\def\d{\dagger}

\begin{document}
\title{Multimode rotationally symmetric bosonic  codes from group-theoretic construction}

\author{Rabsan Galib Ahmed}
\affiliation{Department of Physical Sciences, Indian Institute of Science Education and Research Mohali, Punjab 140306, India}
\affiliation{Wallenberg Centre for Quantum Technology, Department of Microtechnology and Nanoscience, Department of Microtechnology and Nanoscience, Chalmers University of Technology, Göteborg SE-412 96, Sweden}
\affiliation{Institute for Quantum Computing, University of Waterloo and Department of Applied Mathematics, University of Waterloo}
\author{Adithi Udupa}
\affiliation{Wallenberg Centre for Quantum Technology, Department of Microtechnology and Nanoscience, Department of Microtechnology and Nanoscience, Chalmers University of Technology, Göteborg SE-412 96, Sweden}
\author{Giulia Ferrini}
\affiliation{Wallenberg Centre for Quantum Technology, Department of Microtechnology and Nanoscience, Department of Microtechnology and Nanoscience, Chalmers University of Technology, Göteborg SE-412 96, Sweden}

\begin{abstract}
 We introduce a new family of multi-mode, rotationally symmetric bosonic codes inspired by the group-theoretic framework of [Phys. Rev. Lett. 133, 240603 (2024)]. Such a construction inverts the traditional paradigm of code design by identifying codes from the requirement that a group of chosen logical gates should be implemented by means of physically simple logical operations, such as linear optics. Leveraging previously unexplored degrees of freedom within this framework, our construction results in codes that display rotational symmetry across multiple modes, while enabling linear-optics implementation of the full Pauli group. These codes exhibit improved protection against dephasing noise, outperforming both single-mode analogues and earlier multi-mode constructions. Notably, they allow exact correction of correlated dephasing and support qudit encoding in arbitrary dimensions. We analytically construct and numerically benchmark two-mode binomial code instances, and demonstrate that, unlike single-mode rotationally symmetric bosonic codes, these
exhibit no trade-off between protection against dephasing and photon loss. 
\end{abstract}

\maketitle

%Bosonic codes allow for encoding the logical quantum information of a qubit in the (infinitely) many levels of a quantum harmonic oscillator, providing enhanced robustness to noise and holding promise for building scalable architecture of fault-tolerant quantum computers~\cite{albert2025bosonic}. In the recent years, they have been used for enhancing the life time of quantum information beyond one of the elementary experimental components, i.e. to reach the so-called "break-even" regime~\cite{ni2023beating, sivak2023real}. Furthermore, using bosonic codes as the physical qubits allows for reducing the size of discrete-variable codes concatenated to the bosonic code, such as for the surface and LDPC codes, while keeping the same performance ~\cite{PhysRevLett.120.050505,Zhang_2023, Noh-PRA-2020, PRXQuantum.3.010315, PhysRevX.8.021054, Raveendran_2022, PRXQuantum.5.020349, ruiz2025ldpc}  . This in turns induces an increased algorithmic performance as compared to the use of standard qubits~\cite{gouzien2023performance}.
 
Bosonic codes encode a logical qubit into the infinite-dimensional Hilbert space of a quantum harmonic oscillator, enabling protection of quantum information by tailoring the encoding to dominant physical noise processes such as photon loss and dephasing~\cite{albert2025bosonic}. In contrast to conventional qubit-based encodings, which typically require a large number of physical qubits to correct generic errors, bosonic codes can exploit the structure and bias of noise at the hardware level, thereby achieving error suppression with significantly reduced overhead. Recent experiments have demonstrated a prolongation of logical lifetimes beyond those of the underlying physical components, reaching the so-called break-even regime~\cite{ni2023beating, sivak2023real}. Furthermore, concatenating bosonic codes with higher-level quantum error-correcting codes can substantially reduce the resource requirements of fault-tolerant architectures~\cite{PhysRevLett.120.050505,Zhang_2023, Noh-PRA-2020, PRXQuantum.3.010315, PhysRevX.8.021054, Raveendran_2022, PRXQuantum.5.020349, ruiz2025ldpc}. This in turns induces an increased algorithmic performance as compared to the use of standard qubits~\cite{gouzien2023performance}.
However, despite these advantages, important challenges remain, including construction of codes that can tackle a wide range of noise errors and construction of gates that are easily implementable in the existing hardware. Addressing these limitations is essential for assessing the true potential of bosonic codes in scalable fault-tolerant quantum computing.

One way to understand the performance of bosonic codes is in terms of their symmetries. For instance, $N$-order rotationally symmetric bosonic  (RSB) codes~\cite{Grimsmo_2020} have a specific structure of the codewords in Fock space, where only Fock states with multiples of  $N$ are allowed. This structure implies that RSB codes  allow for detecting the loss or gain of up to $N-1$ photons at the same time, while also tolerating phase errors of magnitude up to $\pi/N$.
While such codes have the potential to break even in relevant parameter regimes, finding new bosonic codes with better performance is essential to further reduce the overhead of physical qubits needed in practical implementations of concatenated codes. Furthermore, the implementation of relevant gates such as the logical $X$-gate on such codes is complicated, and requires the use of auxiliary qubits. In particular, for single-mode rotation-symmetric codes, such operations generally require carefully optimized control pulses~\cite{Grimsmo_2020}. Therefore it is essential to design new bosonic codes with good quantum error correcting (QEC) properties, and at the same time with easily implementable sets of logical gates.

While usual approaches to the design of QEC codes consist of first introducing a code, and then finding the operations that implement the logical gates on that code, Ref.~\cite{PhysRevLett.133.240603} takes an opposite approach. It provides a group-theoretic construction where one first selects a chosen group of logical gates, e.g. the Pauli group, to be implemented by simple physical operations, e.g. linear optics, and then a corresponding code is identified, where the desired logical gates are implemented by means of the chosen simple physical operations. However, the construction in Ref.~\cite{PhysRevLett.133.240603} did not take into account general symmetries of the code and their connection to QEC properties. 

In this work, we exploit some previously unexplored freedom in the construction of Ref.~\cite{PhysRevLett.133.240603}, to allow for a more general construction yielding a new family of multi-mode bosonic codes with rotational symmetry, that we therefore call multi-mode RSB codes. These codes retain the advantage of the ones in Ref.~\cite{PhysRevLett.133.240603}, as their logical Pauli operations remain implementable via linear optics, while also offering enhanced performance against dephasing errors compared with the corresponding instances in Ref.\cite{PhysRevLett.133.240603}. Furthermore,  this multimode extension also allows us to encode qudits of arbitrary dimension. We perform a detailed analytical analysis of this family of codes for the qubit case, and also provide numerical results of the performance for a specific instance, namely the dual-rail binomial code, which can be thought as the two-mode generalization of the  binomial code~\cite{Michael2016,Grimsmo_2020}. We provide analytical and numerical evidence showing that, in contrast to the single-mode version of this code, which displays a trade-off between the capability of detecting loss or dephasing errors at increasing order of the discrete rotational symmetry $N$~\cite{Grimsmo_2020}, the  dual-rail binomial code offers a simultaneous improvement of the error detection capability, yielding an overall improvement in the error correction performance over the single-mode case. Furthermore, we show that any two-mode RSB code in our family of codes offers perfect protection against correlated dephasing.

\textit{Preliminaries ---} The essential idea behind the group-theoretic encoding of Ref.~\cite{PhysRevLett.133.240603} is to find a subspace of the Hilbert space of a given physical system, called the \textit{codespace}, that is isomorphic to the logical space, such that a group of logical operations are \textit{easily implementable}. It takes as inputs: a group, $G$, and its representations, $\Lambda$ and $\Pi$, on the logical and physical Hilbert space respectively, where $\Pi$ is constructed using a set of physical operations that we define as easily implementable. The output of this construction is a family of codespaces, from which the codespace performing optimally against relevant noise channels can be chosen. As an application, a two-mode bosonic code has been proposed in Ref.~\cite{PhysRevLett.133.240603}, when $G$ is chosen to be isomorphic to the Pauli group, $\langle X,Z\rangle$ and the physical representation, $\Pi$, is constructed with passive linear operations on two modes. The resulting \textit{Pauli code} has been shown to perform better than the dual-rail code against the pure loss channel in the two modes. However, the construction itself offers more freedom in choosing $G$ and $\Pi$ than it was explored in Ref.~\cite{PhysRevLett.133.240603}. 

\textit{Multimode rotation symmetric bosonic codes ---} We start by showing that an extended  construction allows us for encoding a qudit, having a \textit{d}-dimensional Hilbert space, in a new family of multimode RSB codes, hosted by $d$ bosonic modes. The Pauli group for a qudit is generated by $X$ and $Z$. In the computational basis, their actions are given by $X\ket{k} = \ket{k\oplus 1}, Z\ket{k} = \omega_d^k\ket{k}$, where $\omega_d = e^{2\pi i/d}$. Inspired by, but distinctly from, the  construction of Ref.~\cite{PhysRevLett.133.240603}, we choose the group $G$ to be $G=\langle g,h| h^{2N} = e\rangle$ for an \textit{even} $N$ and the representations $\lambda(g) = X,\lambda(h) = Z$ and~\footnote{Even though the physical representation is given using the passive linear operations on the $d$ modes, note its difference from the one given in~\cite{PhysRevLett.133.240603}, which assumes a physical representation that is isomorphic to the Pauli group itself while the one proposed here is more general.}
\begin{align}
    \pi(g) &= \hat{U}_{BS}\left(\prod_{i=1}^{d-1}e^{-i\frac{\pi}{2}\hat{G}^-_{i,i+1}}\right)^{\dagger}\hat{U}^\dagger_{BS},\label{eqX}\\
    \pi(h) &= \hat{U}_{BS}e^{i\frac{2\pi}{Nd}\hat{a}^\dagger_1\hat{a}_1}\hat{U}^\dagger_{BS}.\label{eqZ}
\end{align}
\begin{figure}
    \centering
    \includegraphics[width=1\linewidth]{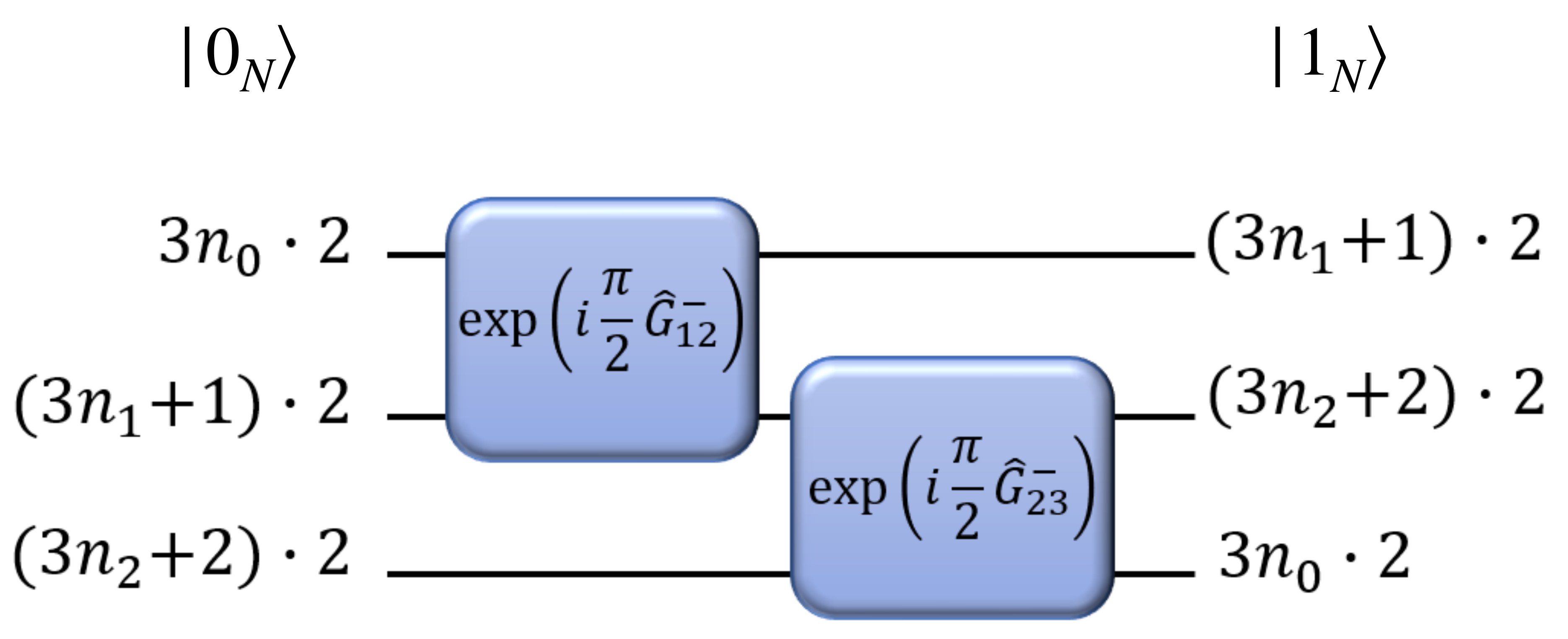}
    \caption{Application of the logical $X$ gate on a qutrit ($d=3$) encoded in a three mode rotation-symmetric bosonic code of order $N=2$. Each of the lines indicates a single bosonic mode.}
    \label{fig:logicalX}
\end{figure}
 Here $\hat{U}_{BS}:=\exp\l\{i\sum_{j,k=1;\\j<k}^d\l(\theta^-_{jk}\hat{G}^{-}_{jk}+\theta^+_{jk}\hat{G}^+_{jk}\r)\r\}$  is a general passive linear operation generated by beam-splitters on all pairs of modes $(j,k)$, with $\hat{G}^{+}_{jk}=\hat{a}_j^\d\hat{a}_k + \hat{a}_k^\d\hat{a}_j$ and $\hat{G}^-_{jk}=i(\hat{a}_j^\d\hat{a}_k-\hat{a}_k^\d\hat{a}_j)$ 
 and $\theta^{\pm}$'s some real numbers parametrising $\hat{U}_{BS}$. Crucially, this mode mixing operation, absent in the construction of Ref.~\cite{PhysRevLett.133.240603}, endows the code with additional freedom in the choice of the encoding modes, which, as we will see, will result in an enhancement of the code performance. A beam-splitter for encoding a multimode bosonic code was considered with NOON states in Ref.\cite{PhysRevA.94.012311}. We find that the $d$-dimensional subspace of $d$ physical modes, where Eqs.(\ref{eqX}) and (\ref{eqZ}) operate as logical $X$ and $Z$ respectively, is spanned by the the computational codewords
 \begin{align}
 \label{eq:codewords-multimode}
     \ket{k_N} = \hat{U}_{BS}\sum_{\{n_i\}=0}^{\infty} f_{n_0\dots n_{d-1}}\bigotimes_{j=k}^{k\oplus ({d-1})}\ket{(dn_{j}+j)N},
 \end{align}
where $\oplus$ is addition modulo $d$, the indexes $\{n_i\}$  run between $0$ and $\infty$, and $f_{n_0\dots n_{d-1}}$'s are coefficients that can be choosen freely, provided they satisfy normalisation. The derivation is given in Ref.~\cite[Sec.S1]{supmat} . We also assume that if the upper limit on the product sign, $\bigotimes$, is less than the lower limit, then the index $j$ runs a period through $\mathbb{Z}/d\mathbb{Z}$. For example, when $k=1,d=3$, $j$ runs through $1,2,0$; hence the product state is $\ket{(3n_1+1)N}\otimes\ket{(3n_2+2)N}\otimes\ket{3n_0N}$. In each of the modes, we have an order-$N$ rotation symmetry, as can be verified by applying the order-$N$ rotation symmetry operator $\hat R^{(k)}_N = e^{i 2 \pi \hat{n}_k/N}$ to the codeword, where $\hat{n}_k = \hat{U}_{BS}\hat{a}_k^\d\hat{a}_k\hat{U}_{BS}^\d$. As an illustration, the action of the logical operator $X$ stemming from Eq.(\ref{eqX}) on a qutrit ($d=3$) encoded in a three mode rotation-symmetric bosonic code of order $N=2$ is illustrated in Fig.\ref{fig:logicalX}.  Note the similarity of the Fock-space structure with the case of single-mode RSB qudits, as defined in Appendix F of Ref. \cite{PhysRevResearch.2.043322}. Distinct multimode bosonic codes with rotational symmetry were introduced from homological rotor codes in Ref. \cite{Yijia2024}. 
Other codes exhibiting rotational symmetry have been studied in the context of single-photon implementations of GKP states \cite{Descamp2024}.

\textit{Two-mode instance and Knill-Laflamme conditions--} As a special case of the multimode encoding of a qudit, we focus on the case when $d=2$: encoding a qubit in two bosonic modes. The most general codewords are explicitly given by
\begin{align}
\label{eq:codewords-general-two-modes-0}
    \ket{0_N} &= \hat{U}_{BS}\sum_{m,n}f_{mn}\ket{2mN,(2n+1)N}\\
    \ket{1_N} &= \hat{U}_{BS}\sum_{m,n}f_{mn}\ket{(2n+1)N,2mN}
   \label{eq:codewords-general-two-modes-1}
\end{align}
with $\hat{U}_{BS} = \exp\l[i\delta(\hat{G}^+_{12}\sin\phi+\hat{G}^-_{12}\cos\phi)\r]$.

By construction, the logical bit-flip gate $X$ acting on these codewords is implemented as 
$\hat{U}_{BS}e^{-i\pi\hat{G}^-_{12}/2}\hat{U}^\dagger_{BS}$,
which corresponds to a SWAP operation. The logical phase-flip gate $Z$ is implemented by $\hat{U}_{BS}e^{i\pi\hat{a}^\dagger_1\hat{a}_1/N}\hat{U}^\dagger_{BS}$
\footnote{We further note that for the codewords defined in Eqs.~(\ref{eq:codewords-general-two-modes-0}) and (\ref{eq:codewords-general-two-modes-1}), when $N$ is odd, the implementation of the $Z$ gate remains unchanged and the logical $X$ gate can instead be implemented as $\hat{U}_{BS}e^{-i\pi\hat{G}^-_{12}/2} e^{i\pi\hat{a}^\dagger_1\hat{a}_1/N}\hat{U}^\dagger_{BS}$}. We focus on loss and dephasing channels.
In the general case of the combined channel of the two, the Kraus representation is given by
\begin{align}
\mathcal{N}_L\circ\mathcal{N}_D\sim\{\hat{L}^{(1)}_p\hat{L}^{(2)}_q\hat{D}^{(1)}_r\hat{D}^{(2)}_s: p,q,r,s\in \mathbb{N}_0\},
\label{eq:Kraus-repr-noise-channel}
\end{align}
where the Kraus operators for the loss and dephasing channel in each mode, $i=1,2$, are respectively given by
\begin{align}
   \hat{L}^{(i)}_p &= \sqrt{\frac{(1-e^{-\kappa_i t})^p}{p!}}e^{-\frac{1}{2}\kappa_i t\; \hat{a}_i^\d\hat{a}_i}\;\hat{a}_i^p,\\
   \hat{D}^{(i)}_r &= \sqrt{\frac{(\gamma_i t)^r}{r!}}e^{-\frac{1}{2}\gamma_i t \;(\hat{a}_i^\d\hat{a}_i)^2}(\hat{a}_i^\d\hat{a}_i)^r.
   \label{eq:Kraus-operators}
\end{align}
The loss and dephasing strength of the channels are respectively quantified by the parameters $\kappa_i t$ and $\gamma_i t$ in each mode. The evolution of the encoded state up to first order under this combined channel is provided in Ref.~\cite[Sec.~S1]{supmat}. 

We then study the Knill–Laflamme (KL) conditions for the above codes for arbitrary coefficients $f_{mn}$ for individual channels.
For a purely loss channel, up to first order in the noise strength, we show in Ref.~\cite[Sec.~S2A]{supmat} that the KL conditions are satisfied provided the coefficients obey $\sum_{m,n}|f_{mn}|^2(m-n) = \tfrac{1}{2}$. For a purely dephasing channel, up to first order in the dephasing noise strength, we see that the KL conditions depend on the encoding angles of the beam-splitter. The KL conditions upto first order in $\gamma_i t$ are satisfied for $N \ge 4$ when $\delta = \pi/4$ (Ref.~\cite[Sec.~S3A]{supmat}).

For both these noise models, in order to correct the effect of the noise channel up to the first order in the noise strength for an arbitrary logical state prepared in the codewords Eqs.\eqref{eq:codewords-general-two-modes-0} and \eqref{eq:codewords-general-two-modes-1}, we propose a two-mode extension of the recovery map outlined for the simplest single-mode bosonic codes \cite{Michael2016,ni2023beating}: a modular number measurement (a generalised number-parity  measurement) indicates the error-syndrome, and corresponding unitary operators are applied. The details of the analytical recovery map for the case of loss and dephasing are provided, respectively, in Ref.~\cite[Secs. S2B and S3B]{supmat}.

\textit{Comparison with other codes--}
We now turn to which families of codes are subsumed in our code family. 
If we disregard the beam-splitter $\hat{U}_{BS}$, choosing $f_{mn} \propto \frac{e^{-2|\alpha|^2}}{\sqrt{(2mN)!((2n+1)N)!}}\alpha^{2mN}\alpha^{(2n+1)N}$ yields the concatenation of a cat code with a dual-rail code~\cite{Knill2001-ap}, distinct from the cat repetition code~\cite{guillaud2019repetition}. This code is also distinct from the pair cat-codes found in ~\cite{Albert_2019}, where the codewords are eigenstate of the difference between the number operators of the two modes. The choice $f_{mn} = 1\;\forall m,n$ yields instead the concatenated dual-rail ideal number-phase code \cite{Grimsmo_2020}, distinct from the repetition number-phase code. For numerical analysis of the performance of our codes, we choose the instance  $f_{mn} = \frac{1}{\sqrt{2}}$ when $(m,n)=(0,0)$ and $(1,0)$, and, $f_{mn} =0$ for any other $(m,n)$. This yields the family of dual-rail binomial code (DRBC)  for $K=2$,
\begin{align}\label{eq:0-N-2N-0logical}
    \ket{0_N} &= \hat{U}_{BS}(\delta,\phi)\frac{1}{\sqrt{2}}(\ket{0}+\ket{2N}) \otimes\ket{N}\\
    \ket{1_N} &= \hat{U}_{BS}(\delta,\phi)\ket{N}\otimes\frac{1}{\sqrt{2}}(\ket{0}+\ket{2N})
    \label{eq:0-N-2N-1logical},
\end{align}
 with $N$ the order of rotation symmetry, and where the encoding-dependent angles $(\delta,\phi)$, here explicited, parameterise the beam-splitting operator $\hat{U}_{BS}$. %We refer to these codes as the \textit{two-mode $K=2$ codes}. 
These codewords are similar, although distinct, to the two-qubit binomial encoded Bell state of Ref. ~\cite{chelluri2025bosonicquantumerrorcorrection} and the two-mode codes of Ref.~\cite{steinbach2000motional} \footnote{Furthermore, our codes are also distinct from the bosonic quantum Fourier codes based on cat states along a similar group-theoretic construction~\cite{Leverrier2026bosonicquantum}.}.
As compared to concatenation with repetition-code, the relatively uniform distribution of the bosonic excitations across the two-modes of our codes holds potential for increased QEC performance.

\textit{Numerical results for DRBC under losses and dephasing channels ---}
We now analyse the performance of the DRBC with $K=2$ defined in Eqs.(\ref{eq:0-N-2N-0logical}) and (\ref{eq:0-N-2N-1logical}), for various values of the beam-splitter parameter. We compare to the "break-even threshold", which is defined here as the average fidelity of a logical qubit constructed with the ground state ($\ket{0}$) and the first excited state ($\ket{1}$) of a harmonic oscillator, called the \textit{trivial encoding}, subjected to relevant noise channels but without a recovery operation \cite{sivak2023real,ni2023beating}. We also compare the performance of our codes with the single-mode binomial code, as well as with the two-mode CLY code of Ref.\cite{Chuang_1997}. Note that the CLY is generally not an instantiation of the family of codes proposed here.
\begin{figure*}
    \centering
    \includegraphics[width=1\linewidth]{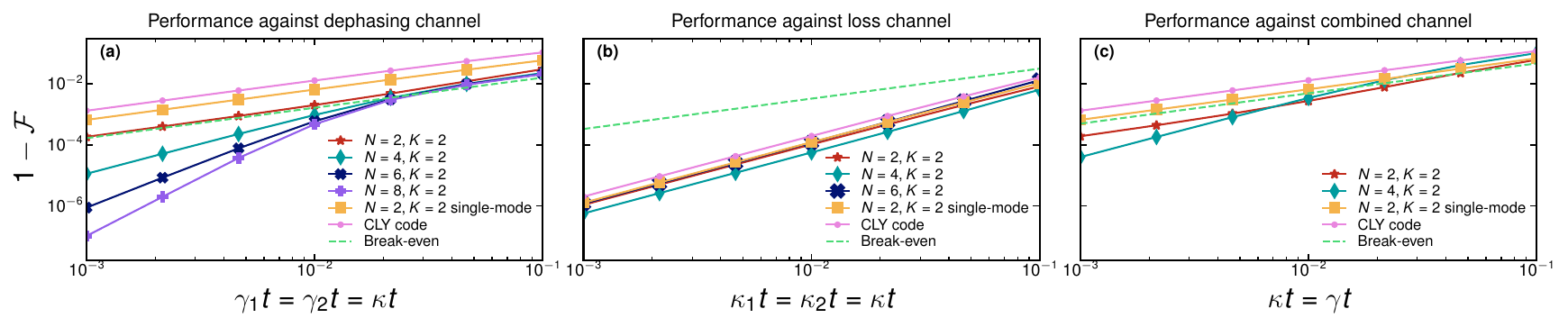}
     \caption{Performance of the simple instances in Eqs.(\ref{eq:0-N-2N-0logical}) and (\ref{eq:0-N-2N-1logical}) of the two-mode rotation symmetric code for the DRBC instances against (a) the dephasing channel with equal rates in both the modes, $\gamma_1 = \gamma_2= \gamma$ , (b) the loss channel with $\kappa_1 = \kappa_2 = \kappa$, and (c) a combination of both of these noisy channels with equal strengths in both the modes and with $\gamma= \kappa$. In all the three cases, we have chosen the optimal values of the beam-splitter angles $(\delta, \phi)$ and we note that under the loss channel, the performance is independent of this choice. We also compare against the $N=2, K=2$ single-mode binomial code, the two-mode CLY code, as well as to the break-even. For dephasing, we also report the performance for $K=2,N=6$ and $K=2,N=8$, corroborating the performance increase at increasing $N$.
     We note that under the dephasing channel the evolution is restricted to a smaller subspace than the full Hilbert space that is required to describe a general evolution of the codewords. This allows us to go to higher values of $N$ to evaluate the performance of the code under the purely dephasing channel. We note that the optimum value of $N$ under the loss channel displayed in panel (b) is around $N=4$, which is similar to the corresponding case of the single-mode code \cite{Grimsmo_2020}. Our two-mode codes show comparable performance to the corresponding $N-$order single-mode code under the loss channel, while they enhance the performance quite significantly against the dephasing channel (panel (a)).}
    \label{fig:performance}
\end{figure*}

In Figure \ref{fig:performance} a), we analyse the performance under a pure dephasing channel. We find with both numerical and analytical analysis (as we will detail further) that the optimal performance of the DRBC with $K=2$ against Gaussian dephasing noise is achieved for the choice of the beam-splitter angles $\delta= \pi/4, 3\pi/4$, $\phi = \pi/(2N)$. The condition $\delta= \pi/4, 3\pi/4$ also corroborates the optimal encoding angles that we found through the KL conditions. For this choice of angles, we see that the performance of the DRBC with $K=2$ is significantly better than for the case of the single-mode binomial code. In Figure.~\ref{fig:performance} b), we study the performance under pure loss channel, and we see that the DRBC behave very similar to the single-mode case while still doing better than CLY code.  

For single-mode RSB codes, there is a trade-off~\cite{Grimsmo_2020}:  while the performance against losses increases at increasing $N$ (up to an optimal value of $N$),  for dephasing it decreases at increasing $N$. After this optimal value of $N$, the performance decreases with $N$ for both noise channels. Strikingly, we observe instead a significant improvement in performance under the dephasing noise channel of our DRBC with $K=2$ for the optimal choice of the encoding angles $(\delta,\phi)$, which results in  a simultaneous improvement of the performance with increasing $N$ up to an optimal value of $N$, see Figure \ref{fig:performance} a) and b). In Fig. \ref{fig:performance} c) we present the results for the combined dephasing-loss channel. The performance analysis for each of the noise models studied has been carried out by numerically solving the semi-definite programming (SDP) of finding the optimal recovery map \cite{Fletcher_2007}.  Evaluating the performance for larger values of $N$ and $K$ using semidefinite programming becomes computationally challenging due to the rapidly increasing dimension of the Hilbert space. We provide numerical results on near-optimal fidelity that supports and extends the results obtained from SDP in Ref.~\cite[Sec. 4]{supmat}.

{\it Understanding the code perfomance against dephasing in terms of Knill-Laflamme conditions and codewords distinguishability ---}
For dephasing noise using the continuous Kraus representation, we show in Ref.~\cite[Sec.S5A]{supmat} that, for optimal angles $\delta= \pi/4, 3\pi/4$, $\phi = \pi/(2N)$,  as $N$ approaches infinity, the KL conditions for dual codewords get closer to being exactly satisfied at any order of dephasing parameter. This validates the enhancement of performance at increasing $N$ observed numerically. 
This increased performance as compared to the case of single-mode RSB codes can be further analysed in terms of the codewords distinguishability under phase measurements. 
In a previous study \cite{udupa2025performance}, it was  found that the performance of the single-mode rotation symmetric codes for the Knill teleportation based error-correcting (telecorrection) circuit against a purely dephasing noise is upper bounded by a monotically increasing function of the distinguishability between the dual codewords, $\ket{\pm_N}$. In order to assess  this distinguishability, one can employ the \textit{canonical phase measurement} \cite{Leonhardt_95,Grimsmo_2020}. What was essentially found in \cite{udupa2025performance} is that, as long as the probability distributions of the outcomes of the phase measurements for the codewords, $\ket{+_N}$ and $\ket{-_N}$, do not significantly overlap under the evolution of a dephasing noise, it is possible to recover the logical state faithfully on average. 
Here we extend this understanding to the case of two-mode rotation-symmetric codes by analysing the phase measurements on the two modes. Unlike the case of single-mode codes, where the phase measurement outcomes lie in a circle, $[0,2\pi)\cong S^1$, those in the case of two-mode codes lie in a torus, $S^1\times S^1$. Therefore, in contrast to the single-mode codes where the distributions inevitably overlap in the circle, $S^1$, for the two-mode codes, with different choices of the encoding angles, $\delta,\phi$, the distributions can be made to avoid each other on the torus, $S^1\times S^1$ (see Ref.~\cite[Sec.S5B]{supmat} for more details). We emphasize that, despite these arguments are based on distinguishability with respect to phase measurements, for correcting the dephasing noise to the leading orders for our two-mode code we use the modular number measurement to reveal the error syndrome as described in Ref.~\cite[Sec.S2B and S3B]{supmat}.  

 We have also verified that the increase in performance of the DRBC for the optimal choice of the encoding angles, as compared to the single-mode binomial code,  is robust even in the case of more sophisticated instances of dephasing channels, such as non-Markovian dephasing induced by random telegraph noise~\cite{decohrtn1, decohrtn2, decohrtn3, decohrtn4, benedetti2014non, udupa2025performance}, see Ref.~\cite[Sec.S6]{supmat}.

\textit{Performance of two-mode RSB under correlated dephasing ---} 
Correlated dephasing can arise e.g. as a result of the
usually strong dephasing of flux tunable couplers such as SNAIL-based modes \cite{lescanne2020exponential,PRXQuantum.4.020355}, noise which is picked up by the coupled bosonic modes. This can be modelled through a stochastic interaction Hamiltonian, $\hat{H}(t) = \nu c(t)(\hat{a}_1^\dagger\hat{a}_1+\hat{a}_2^\dagger\hat{a}_2)$, where $c(t)$ is a stochastic variable. The effective noise channel is given in this case by
\begin{align}  
	\mathcal{N}(\hat{\rho}) = \int_{-\infty}^\infty d\phi \;p_t(\phi)e^{i\phi(\hat{n}_1+\hat{n}_2)}\;\hat{\rho}\;e^{-i\phi(\hat{n}_1+\hat{n}_2)}.
    \label{eq:corr_evol}
\end{align}

Here we show that for any choice of two-mode RSB code in Eqs.(\ref{eq:codewords-general-two-modes-0}) and (\ref{eq:codewords-general-two-modes-1}),
the quantum circuit in Fig.~\ref{fig:corr-deph}  corrects exactly any such arbitrary correlated dephasing noise of the form in Eq.(\ref{eq:corr_evol}). Here, modes $1,2$ and $3,4$ encode the control and the target qubits, respectively. The first controlled-$X$ ($CX$) gate in Fig.~\ref{fig:corr-deph} between two qubits encoded in order-$N$ (control qubit) and order-$M$ (target qubit) two-mode RSB code is implemented by the unitary $CX_{NL} = \exp\l(i\frac{\pi}{2N}\hat{n}_1\otimes\hat{G}^{-}_{34}\r)$.  The second control unitary in Fig.~\ref{fig:corr-deph} is given by $\exp\l(i\frac{\pi}{2N}\hat{n}_3\otimes\hat{G}^{-}_{12}\r)$. %The two $CX$ gates effectively act as a SWAP gate, since the auxuliary mode is prepared in the logical state $|0\rangle$, so that the noise is effectively swapped to the auxiliary mode after the circuit.
The two $CX$ gates act as a SWAP gate, since the auxuliary mode is prepared in the logical state $|0\rangle$, and the correlated dephasing noise commutes through the $CX$ gates, effectively swapping the noise-free state in the data rail to the auxiliary mode rail. 
In Ref.~\cite[Sec.S7]{supmat}, we provide a more detail  derivation of the action of this circuit.

\begin{figure}
    \centering
    \includegraphics[width=1\linewidth]{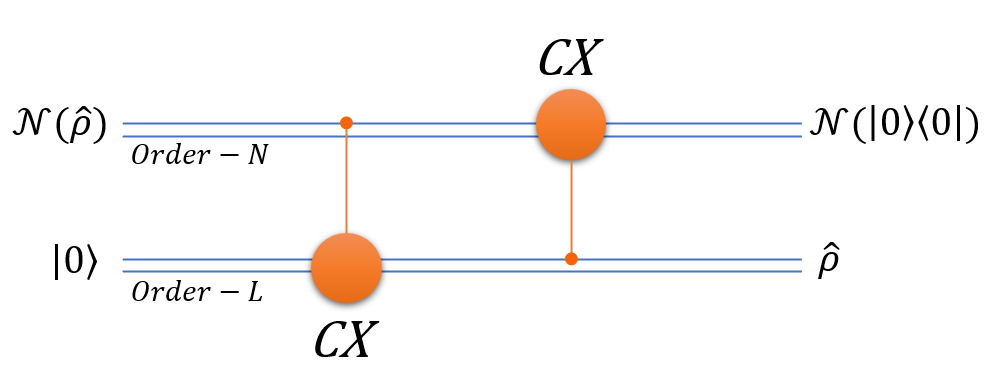}
    \caption{Exact error-correcting circuit for correlated stochastic dephasing noise.}
    \label{fig:corr-deph}
\end{figure}

\textit{Application of other gates ---}
The group of logical operators that we have chosen to be implemented through the \textit{easy} operations, namely the passive linear operations, is the Pauli group. However, that is not enough to achieve universal quantum compution. For that purpose we additionally give a prescription for the application of the Hadamard gate ($H$), the phase gate ($S$), the $T$-gate ($T$) and the two-qubit control-Z gate ($CZ$). 
Similarly to the case of the Pauli code discussed in  Ref.~\cite{PhysRevLett.133.240603}, for our family of codes, for general even $N$, the $S$ and the $CZ$ gate can be applied using the non-linear interactions of a microwave resonator~\cite{Grimsmo_2020}. Explicitly, for an even-$N$ two-mode code, we have $\hat{S}_N = \exp(i\frac{\pi}{2N^2}\hat{n}_1^2)$, implementable using the self-Kerr interaction. The $CZ$ gate between two qubits encoded in order-$N$ and order-$M$ can be applied similarly using the cross-Kerr interaction between two microwave resonator: $CZ_{NM} = \exp(i\frac{\pi}{NM}\hat{n}_1\otimes\hat{n}_3)$~\cite{Grimsmo_2020}, where the interaction happens between the first mode of each of the two encoded bosonic qubits. The $H$ and $T$ gates can be implemented using teleported gates, similar to what was proposed in~\cite{Grimsmo_2020}. For this purpose, an auxiliary qubit, encoded in two modes, is initially prepared in the states, $\ket{+_N}=\frac{1}{\sqrt{2}}(\ket{0_N}+\ket{1_N})$ and $\ket{T_N} = \frac{1}{\sqrt{2}}(\ket{0_N}+e^{i\pi/4}\ket{1_N})$ respectively, followed by a coupling to the two-mode data qubit using a $CZ$ gate and a phase measurement on the data qubit. The resultant state of the auxiliary qubit contains the desired action of the gates up to an action of a logical $X$. The circuits are included in Ref.~\cite[Sec.S8]{supmat}. 

\textit{Conclusive remarks and open questions ---}
We have introduced a new family of multimode bosonic codes, which extends the rotationally symmetric bosonic codes introduced in Ref.~\cite{Grimsmo_2020} to the multimode, multidimensional case, by endowing the group theoretic construction  introduced in Ref.~\cite{PhysRevLett.133.240603} with the addition of beam-splitters mixing the codewords. An analysis of the two-mode dual-rail binomial instance for $K=2$ reveals increased performance as compared to the single-mode case, in particular against dephasing noise, where the trade-off between resilience towards dephasing and loss errors at increasing discrete symmetry order $N$ that affects the single-mode RSB codes (up to the optimum $N$) is resolved. This increase in performance achieved by increasing the number of modes in the codes echoes similar results found for translationally symmetric bosonic codes~\cite{PRXQuantum.3.010335, Nordquantique}, but was not explored yet for RSB codes, to the best of our knowledge. 

Several directions for further research arise from our work. For example, the very general codewords presented in Eq.(\ref{eq:codewords-multimode}) yield a family of codes, with vast space for numerical optimization of the parameters, given a specific noise model, such as qudit dimensionality, order of rotational symmetry, and coefficients of the superposition. Furthermore, for the codes studied in this work we have chosen the Pauli group as the group to be implemented covariantly; it is an interesting question to ask what codes would stem by selecting other representations of the Clifford group or of groups homomorphically equivalent to it, than those considered in Ref.~\cite{PhysRevLett.133.240603}.
Finally, it is interesting to consider the case where physically implementable operations are not constrained to passive linear optical operations, but correspond instead to more sophisticated operations, such as those implementable with superconducting circuits~\cite{hillmann2020universal,eriksson2024universal}. 

\textit{Acknowledgment ---} 
We acknowledge useful discussions with Tahereh Abad, Victor Albert, Axel Eriksson,  Yvonne Gao, Simone Gasparinetti, Mats Granath, Trond Haug, Kunal Dhanraj Helambe, Maryam Khanahmadi, Martin Jirelow, Anthony Leverrier, Anja Metelmann, Lukas Splitthof, Peter van Loock, and Suocheng Zhao. We acknowledge Alberto Salvador for the numerical calculations on near-optimal fidelity results reported in the supplementary material file, as well as Julius Andersson, Carl Svensson, Hang Zhou and Griffin Hiscoke for having derived the Knill-Laflamme conditions for the case of losses to second order in the loss rate. We thank Debjyoti Biswas and Nikhil Sharma for useful cross validation of  Knill-Laflamme conditions of loss channel.
G.F.\ acknowledges funding from the European Union’s Horizon Europe Framework Programme (EIC Pathfinder Challenge project Veriqub) under Grant Agreement No.\ 101114899, the Olle Engkvist Foundation, and the Swedish Research Council through the project grant VR DAIQUIRI. G.F. and A.U. acknowledge support from the Knut and Alice Wallenberg Foundation through the Wallenberg Center for Quantum Technology (WACQT). G.F., A.U. and R. G. A. acknowledge funding from Chalmers Area of Advance Nano, as well as resources at the Chalmers Centre for Computational Science and Engineering (C3SE).

\bibliographystyle{apsrev4-1}
\bibliography{bibliography}

\clearpage
\appendix
\begin{widetext}

\title{Supplemental Material for “Multimode rotationally symmetric bosonic  codes from group-theoretic construction”}

\author{Rabsan Galib Ahmed}
\affiliation{Department of Physical Sciences, Indian Institute of Science Education and Research Mohali, Punjab 140306, India}
\affiliation{Department of Microtechnology and Nanoscience, Chalmers University of Technology, Göteborg SE-412 96, Sweden}
\affiliation{Institute for Quantum Computing, University of Waterloo and Department of Applied Mathematics, University of Waterloo}
\author{Adithi Udupa}
\affiliation{Department of Microtechnology and Nanoscience, Chalmers University of Technology, Göteborg SE-412 96, Sweden}
\author{Giulia Ferrini}
\affiliation{Department of Microtechnology and Nanoscience, Chalmers University of Technology, Göteborg SE-412 96, Sweden}

\maketitle
% Optional: to restart numbering separately from main text
\setcounter{equation}{0}
\renewcommand{\theequation}{S\arabic{equation}}
\setcounter{figure}{0}
\renewcommand{\thefigure}{S\arabic{figure}}
\setcounter{table}{0}
\renewcommand{\thetable}{S\arabic{table}}

\section{Multimode rotation-symmetric codewords}\label{appsec:multimode}

In this section, we provide a derivation of the qudit codewords Eq.(3). Firstly, note that unlike~\cite{PhysRevLett.133.240603} 
where both the group $G$ and the  physical representation $\Pi$ are isomorphic to the Pauli group, we have utilised the fact that it is sufficient for there to exist a homomorphism, $\Lambda$, from $G$ to the Pauli group and a physical representation, $\Pi$, that acts as $\Lambda$ on at least one two-dimensional subspace.

For the derivation it suffices to work out the case for $\hat{U}_{BS}=\mathbb{I}$. We start by considering the general superposition of all possible states supported by $d$ modes, given by
\begin{align}
    \ket{k_N} = \sum_{\{m_i\}=0}^{\infty}f^{(k)}_{m_0,\dots ,m_{d-1}}\ket{m_0,m_1,\dots,m_{d-1}}.
\end{align}
By imposing that the physical representation of the group element $h$ acts as in Eq.(2), i.e. that $\Pi(h)\ket{k_N} = e^{i\frac{2\pi}{Nd}\hat{a}^\dagger_1\hat{a}_1}\ket{k_N}=\omega_d^k\ket{k_N}$, with $\omega_d=e^{2\pi i/d}$, we find,

\begin{align}
\label{eq:pi(h)_cond}
   f^{(k)}_{m_0,\dots, m_{d-1}} &= 0, \quad \text{unless } 
   e^{i\frac{2\pi}{Nd}m_0} = e^{i\frac{2\pi k}{d}} \nonumber\\
   \implies~& m_0 =  (n_0 d + k) N,~ \forall n_0 \in \mathbb{N}_0.
\end{align}

By imposing the physical representation of group element $g$, we require that $\Pi(g)\ket{k_N} = \ket{(k\oplus 1)_N}$. Using the relation, $e^{i\frac{\pi}{2}\hat{G}^-_{i,i+1}} \hat{a}^{\dagger}_i e^{-i\frac{\pi}{2}\hat{G}^-_{i,i+1}} = -\hat{a}^{\dagger}_{i+1}$, we obtain
\begin{align}
    f^{(k)}_{m_0,\dots, m_{d-1}} (-1)^{\sum_{i=1}^{d-1}m_i} = f^{(k\oplus 1)}_{m_1,\dots, m_{d-1},m_0}.
\end{align}
From Eq.~(\ref{eq:pi(h)_cond}), we also see that 
$f^{(k\oplus 1)}_{m_1,\dots, m_{d-1},m_0} = 0, \quad \text{unless } 
m_1 = \left[n_1 d + (k \oplus 1)\right] N.$ 
This implies
\begin{equation}
    f^{(k)}_{m_0,\dots, m_{d-1}} = 0, \quad \text{unless } 
    m_1 = \left[n_1 d + (k \oplus 1)\right] N.
\end{equation}
Similarly, applying $\Pi(g)$ on $\ket{k\oplus 1}$, we arrive at the condition $f^{(k\oplus 1)}_{m_1,\dots, m_{0}} (-1)^{\sum_{i=1}^{d-1}m_i} = f^{(k\oplus 2)}_{m_2,\dots, m_{0},m_1}$ where we again use Eq.~(\ref{eq:pi(h)_cond}) to obtain the fact that $f^{(k\oplus2)}_{m_2, \dots, m_1}$, and thus $f^{(k)}_{m_0,\dots, m_{d-1}}$ is equal to 0, unless $m_2 = \l[n_2d+(k\oplus 2)\r]N$.

By induction, we therefore have
\begin{align}
f^{(k)}_{m_0,\dots,m_j,\dots ,m_{d-1}} = 0,\text{ unless }m_j = \l[n_jd+(k\oplus j)\r]N
\end{align}
for all $j=0,\dots,d-1$. Furthermore, since $N$ is an even number, $(-1)^{\sum_{i=1}^{d-1}m_i} = 1$ for all the fock-states that have non-zero amplitudes in the codewords. The codewords thus take the form given in Eq.(3). 

The set-up and the error correction procedure that we investigate in this paper is given as follows (see Figure~\ref{fig:circuit}). First, we prepare the logical state $\hat{\rho}$ in the two-mode encoding in the modes through which noise occurs ($\hat{a}_1,\hat{a}_2$). Hence,
\begin{align}
	\bar{\hat{\rho}} = \sum_{i,j=0}^1\rho_{ij}\ket{\bar{i}_N}\bra{\bar{j}_N},
\end{align}
where the bar denotes the logical codewords in non-transformed basis, i.e. before the beam-splitter operation. 
 Then we apply the beam-splitter action, 
 \begin{equation}
 \label{eq:beam-splitter-explicit}
 \hat{U}_{BS} = \exp\l[i\delta(\hat{G}^+_{12}\sin\phi+\hat{G}^-_{12}\cos\phi)\r] = \exp\l[ \delta(\hat a_2^\dagger \hat a_1 e^{i \phi} - \hat a_1^\dagger \hat a_2 e^{-i \phi})\r],
 \end{equation}
 and let the system go through the noise channel $\mathcal{N} = \mathcal{N}_L\circ\mathcal{N}_D$, and then apply the inverse beam-splitter action $\hat{U}_{BS}^\d$ before continuing with rest of the recovery operation, $\tilde{\mathcal{R}}$, where we have  $\mathcal{R}(\bar{\hat{\rho}}) = \tilde{\mathcal{R}}(\hat{U}_{BS}^\dagger\;\bar{\hat{\rho}}\;\hat{U}_{BS})$. %\commg{$=\tilde{\mathcal{R}}({\hat{\rho}})$? lets discuss this notation together}
 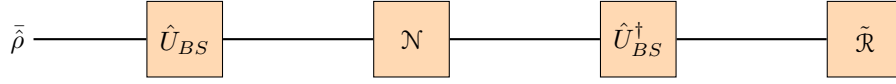
\begin{figure}[ht!]
	\centering
	\begin{tikzpicture}[node distance=3cm]
		\coordinate [label=left:$\bar{\hat{\rho}}$] (A) at (-2,0){};
		\node (pro1) [process2] {$\hat{U}_{BS}$};
		\node (pro2) [process2, right of = pro1] {$\mathcal{N}$};
		\node (pro3) [process2, right of = pro2] {$\hat{U}^\d_{BS}$};
		\node (rec) [process2, right of = pro3] {$\tilde{\mathcal{R}}$};
		\draw [thick] (A)--(pro1);
		\draw [thick] (pro1)--(pro2);
		\draw [thick] (pro2)--(pro3);
		\draw [thick] (pro3)--(rec);
	\end{tikzpicture}
	\caption{An encoding and error-correcting procedure for the rotation-symmetric codes on two modes. }      
	\label{fig:circuit}
\end{figure}
The beam-splitter transforms the two modes $\hat{a}_1$ and $\hat{a}_2$ into $\hat{b}_1$ and $\hat{b}_2$ as follows
\begin{align}
\label{eq:beam-splitter-explicit2}
	\hat{b}_1 &= \hat{U}^\d_{BS}\hat{a}_1\hat{U}_{BS} = \hat{a}_1\cos\delta +\hat{a}_2e^{-i\phi}\sin\delta \nonumber \\
	\hat{b}_2 &= \hat{U}^\d_{BS}\hat{a}_2\hat{U}_{BS} = \hat{a}_2\cos\delta -\hat{a}_1e^{i\phi}\sin\delta.
\end{align}

 With the Kraus representation provided in Eq.(8), the evolution of a state under these channels can be written as  
\begin{align}\label{eq:noise-two-mode}
	\tilde{\mathcal{N}}(\bar{\hat{\rho}}) =&\hat{U}_{BS}^\d \mathcal{N}(\hat{U}_{BS} \bar{\hat{\rho}} \hat{U}_{BS}^\d) \hat{U}_{BS} \nonumber \\
    = &\sum_{p,q,r,s=0}^\infty \hat{U}_{BS}^\d\hat{L}^{(1)}_p\hat{L}^{(2)}_q\hat{D}^{(1)}_r\hat{D}^{(2)}_s\hat{U}_{BS}\; \bar{\hat{\rho}}\; \hat{U}_{BS}^\d\hat{D}^{(2)\d}_s\hat{D}^{(1)\d}_r\hat{L}^{(2)\d}_q\hat{L}^{(1)\d}_p\hat{U}_{BS}.
\end{align} 
Note that the modified noise channel under the beam-splitter operation is denoted by  $\tilde{\mathcal{N}}$ instead of $\mathcal{N}$. 
%-----
We can further re-write
\begin{align}
\hat{U}_{BS}^\d\hat{L}^{(1)}_p\hat{L}^{(2)}_q\hat{D}^{(1)}_r\hat{D}^{(2)}_s\hat{U}_{BS} = \hat{\tilde{L}}^{(1)}_p\hat{\tilde{L}}^{(2)}_q\hat{\tilde{D}}^{(1)}_r\hat{\tilde{D}}^{(2)}_s,
\end{align}
where, using Eq.(10), the new Kraus operators are
\begin{align}
	\hat{\tilde{L}}^{(1)}_p = \sqrt{\frac{(1-e^{-\kappa_1 t})^p}{p!}}e^{-\frac{1}{2}\kappa_1 t\; \hat{b}_1^\d\hat{b}_1}\;\hat{b}_1^p\hspace{2cm}
	\hat{\tilde{L}}^{(2)}_q = \sqrt{\frac{(1-e^{-\kappa_2 t})^q}{q!}}e^{-\frac{1}{2}\kappa_2 t\; \hat{b}_2^\d\hat{b}_2}\;\hat{b}_2^q \label{eq:loss_kraus_new}\\
	\hat{\tilde{D}}^{(1)}_r =\sqrt{\frac{(\gamma_1 t)^r}{r!}}e^{-\frac{1}{2}\gamma_1 t \;(\hat{b}_1^\d\hat{b}_1)^2}(\hat{b}_1^\d\hat{b}_1)^r\hspace{2cm}
	\hat{\tilde{D}}^{(2)}_s =\sqrt{\frac{(\gamma_2 t)^s}{s!}}e^{-\frac{1}{2}\gamma_2 t \;(\hat{b}_2^\d\hat{b}_2)^2}(\hat{b}_2^\d\hat{b}_2)^s.
    \label{eq:dephasing_kraus_new}
\end{align}

The recovery procedure involves a non-destructive modular number measurement in the ($\hat{a}_1, \hat{a}_2$) basis to detect the error syndrome and a corresponding unitary operator is applied. This modular number measurement in each of the modes is defined by the following set of POVMs:  $\{\hat{\mathcal{P}}_{pq}=\hat{\mathcal{P}}_p\otimes\hat{\mathcal{P}}_q \}$ where $\hat{\mathcal{P}}_p := \sum_{m=0}^\infty \ketbra{mN+p}$~\cite{Grimsmo_2020, Michael2016}. 
We intend to find the recovery map that corrects the error induced by the noise channel up to the first order in the noise strengths, $\kappa_i t$ and  $\gamma_i t$ corresponding to the loss and dephasing channel respectively.
Up to the first order in $\kappa_i t$ and $\gamma_i t$, the effect of the noise channel can be written as
\begin{align}
	\tilde{\mathcal{N}}(\hat{\rho}) \approx \hat{\rho} -\frac{\kappa_1 t}{2}\{\hat{b}_1^\d\hat{b}_1,\hat{\rho}\}-\frac{\kappa_2 t}{2}\{\hat{b}_2^\d\hat{b}_2,\hat{\rho}\}-\frac{\gamma_1 t}{2}\{(\hat{b}_1^\d\hat{b}_1)^2,\hat{\rho}\}-\frac{\gamma_2 t}{2}\{(\hat{b}_2^\d\hat{b}_2)^2,\hat{\rho}\}\nn\\	
	+\kappa_1 t\;\hat{b}_1\hat{\rho}\hat{b}_1^\d+\kappa_2 t\;\hat{b}_2\hat{\rho}\hat{b}_2^\d +\gamma_1 t\;(\hat{b}_1^\d\hat{b}_1)\hat{\rho}(\hat{b}_1^\d\hat{b}_1)+\gamma_2 t\;(\hat{b}_2^\d\hat{b}_2)\hat{\rho}(\hat{b}_2^\d\hat{b}_2).
    \label{eq:noise_channel_state_evolve}
\end{align}

Based on this expansion, we give recovery procedures in the presence of purely loss and purely dephasing channels in Secs. \ref{appsec:recovery_loss} and \ref{appsec:recovery_deph} respectively, for general codewords of our two-mode codes, corresponding to general $f_{mn}$ in Eqs.(4) and (5).

\section{Recovery of the two-mode RSB code against the loss channel up to first order in noise strength}
\label{appsec:recovery_loss}

We rewrite the codewords given in Eq.~(4) and (5) of the main text in the non-transformed basis, where the mixing operation $\hat{U}_{BS}(\delta,\phi)$ has been absorbed into the error channel, as 
\begin{align}
    \ket{\bar{0}_N} &= \sum_{m,n}f_{mn }\ket{2mN,(2n+1)N}\\
    \ket{\bar{1}_N} &= \sum_{m,n}f_{mn}\ket{(2n+1)N,2mN}.
     \label{eq:non-rotated_codewords}
\end{align}

\subsection{Knill-Laflamme conditions for purely loss channel}

We now check the Knill-Laflamme (KL) conditions for the loss channel up to first order in $\kappa_i t$. From Eq.~(\ref{eq:loss_kraus_new}), the Kraus operators to first order in $\kappa_i t$ are
\begin{align}
	\hat{\tilde{L}}^{\prime(i)}_0= (\mathbb{I} - \frac{1}{2}\kappa_i t \hat{b}^{\dagger}_i\hat{b}_i)\hspace{2cm}
\hat{\tilde{L}}^{\prime(i)}_1= \sqrt{\kappa_i t} \hat{b}_i.
\label{eq:loss_kraus_first_order}
\end{align}

The corresponding error operators are
\begin{align}
    \hat{E}_1 &= \hat{\tilde{L}}^{\prime(1)}_0  \hat{\tilde{L}}^{\prime(2)}_0 \\
    \hat{E}_2 &= \hat{\tilde{L}}^{\prime(1)}_0  \hat{\tilde{L}}^{\prime(2)}_1 \\
    \hat{E}_3 &= \hat{\tilde{L}}^{\prime(1)}_1  \hat{\tilde{L}}^{\prime(2)}_0 \\
    \hat{E}_4 &= \hat{\tilde{L}}^{\prime(1)}_1  \hat{\tilde{L}}^{\prime(2)}_1.
\end{align}

The KL conditions require
\begin{equation}
    \bra{\bar{i}_N} \hat{E}^{\dagger}_a \hat{E}_b \ket{\bar{j}_N} = C_{ab} \delta_{ij},
\end{equation}
where $a,b=1,2,3,4$ and $i,j=0,1$. Among the ten possible combinations of $\hat{E}_a^\dagger \hat{E}_b$, three appear only at order higher than $(\kappa_i t)$ and can therefore be neglected. We evaluate the remaining seven cases below.

\textbf{Case 1:} $\hat{E}_1^{\dagger}\hat{E}_1$
\begin{align}
    \hat{E}_1^{\dagger} \hat{E_1}&= (\mathbb{I}- \frac{1}{2}\kappa_1 t \hat{b}^{\dagger}_1\hat{b}_1)(\mathbb{I}- \frac{1}{2}\kappa_2 t \hat{b}^{\dagger}_2\hat{b}_2)(\mathbb{I}- \frac{1}{2}\kappa_1 t \hat{b}^{\dagger}_1\hat{b}_1)(\mathbb{I}- \frac{1}{2}\kappa_2 t \hat{b}^{\dagger}_2\hat{b}_2) \\
    &= -\kappa_1 t \hat{b}_1^\dagger \hat{b}_1-\kappa_2 t \hat{b}_2^\dagger \hat{b}_2 + 
    \mathcal{O}((\kappa_i t)^2)\\
    &= -\kappa_1 t (\cos^2\delta~ \hat{a}_1^{\dagger}\hat{a}_1 +\sin^2~\delta \hat{a}_2^{\dagger}~\hat{a}_2+ \sin \delta \cos\theta(e^{i\phi}\hat{a}_1^{\dagger}a_2 +  e^{-i\phi}\hat{a}_2^{\dagger}a_1) )\\
    &~~~~-\kappa_2 t (\cos^2\delta ~\hat{a}_2^{\dagger}\hat{a}_2+ \sin^2 \delta \hat{a}_1^{\dagger}~\hat{a}_1+ \sin \delta \cos\theta(-e^{i\phi}\hat{a}_1^{\dagger}a_2 +  e^{-i\phi}\hat{a}_2^{\dagger}a_1)).
\end{align}
For the codewords in Eq.~(\ref{eq:non-rotated_codewords}), the off-diagonal matrix element is proportional to $\delta_{(2l+1)N,~ 2kN}\delta_{2mN+1, (2n+1)N-1}$ which is zero for $N\ge2$ as required by KL conditions to be satisfied. Although we only consider even $N$, we note here that if $N=1$, this matrix element will not vanish and thus not satisfying KL condition. Evaluating the diagonal terms yields
\begin{align}
    \bra{\bar{0}_N}\hat{E}_1^{\dagger} \hat{E_1}\ket{\bar{0}_N}=& -\kappa_1 t (\cos^2 \delta \sum_{m,n}|f_{mn}|^2 2mN + \sin^2 \delta \sum_{m,n} |f_{mn}|^2 (2n+1)N) \\ \nonumber
    &-\kappa_2 t (\cos^2 \delta \sum_{m,n}|f_{mn}|^2 (2n+1)N + \sin^2 \delta \sum_{m,n} |f_{mn}|^2 2mN) \\
    \bra{\bar{1}_N}\hat{E}_1^{\dagger} \hat{E_1}\ket{\bar{1}_N} =& -\kappa_1 t(\cos^2 \delta \sum_{m,n}|f_{mn}|^2 (2n+1)N + \sin^2 \delta \sum_{m,n} |f_{mn}|^2 2mN) \\ \nonumber
    &-\kappa_2 t  (\cos^2 \delta \sum_{m,n}|f_{mn}|^2 2mN + \sin^2 \delta \sum_{m,n} |f_{mn}|^2 (2n+1)N).
\end{align}

This leads to the condition
\begin{equation}
    \cos^2 \delta \sum_{m,n}|f_{mn}|^2 (2mN- (2n+1)N) = \sin^2 \delta \sum_{m,n} |f_{mn}|^2 (2mN-(2n+1)N).
    \label{eq:loss_delta_cond}
\end{equation}

\textbf{Case 2:} 
$\hat{E}_2^{\dagger}\hat{E}_1$
\begin{align}
\hat{E}_2^{\dagger}\hat{E}_1 =& \sqrt{\kappa_2 t} \hat{b}_2^{\dagger} (\mathbb{I}- \frac{1}{2}\kappa_1 t \hat{b}^{\dagger}_1\hat{b}_1)(\mathbb{I}- \frac{1}{2}\kappa_1 t \hat{b}^{\dagger}_1\hat{b}_1)(\mathbb{I}- \frac{1}{2}\kappa_2 t \hat{b}^{\dagger}_2\hat{b}_2)\\
=&\sqrt{\kappa_2 t} \hat{b}_2^{\dagger} +\mathcal{O}((\kappa_i t)^2)\\
=& \sqrt{\kappa_2 t} (\hat{a}_1^{\dagger}\cos\delta - e^{-i\phi}\hat{a}_2^{\dagger}\sin\delta).
\end{align}\\

Since $\bra{\bar{i}_N}\hat{a}_{1,2}^{\dagger}\ket{\bar{j}_N}=0$ for all $i,j$, the respective KL conditions are satisfied.\\

\textbf{Case 3:} $\hat{E}_3^{\dagger}\hat{E}_1$
\begin{align}
\hat{E}_3^{\dagger}\hat{E}_1 =&  (\mathbb{I}- \frac{1}{2}\kappa_1 t \hat{b}^{\dagger}_2\hat{b}_2)\sqrt{\kappa_1 t}\hat{b}_1^{\dagger} (\mathbb{I}- \frac{1}{2}\kappa_1 t \hat{b}^{\dagger}_1\hat{b}_1)(\mathbb{I}- \frac{1}{2}\kappa_2 t \hat{b}^{\dagger}_2\hat{b}_2)\\
=&\sqrt{\kappa_1 t} \hat{b}_1^{\dagger} +\mathcal{O}((\kappa_i t)^2).
\end{align}

This produces the same operator structure as the previous case.\\

\textbf{Case 4:} $\hat{E}_4^{\dagger}\hat{E}_1$
\begin{align} 
\hat{E}_4^{\dagger}\hat{E}_1 &= \sqrt{\kappa_1 t}\sqrt{\kappa_2 t} \hat{b}_2^{\dagger}\hat{b}_1^{\dagger}(\mathbb{I}- \frac{1}{2}\kappa_1 t \hat{b}^{\dagger}_1\hat{b}_1)(\mathbb{I}- \frac{1}{2}\kappa_2 t \hat{b}^{\dagger}_2\hat{b}_2) \nonumber \\
&= \sqrt{\kappa_1 t}\sqrt{\kappa_2 t}\hat{b}_2^{\dagger}\hat{b}_1^{\dagger}+\mathcal{O}((\kappa_i t)^2) \nonumber \\
&= \sqrt{\kappa_1 t}\sqrt{\kappa_2 t} (\hat{a}_1^\d \hat{a}_2^\d (\cos^2\delta - \sin^2\delta) + (\hat{a}_2^\d \hat{a}_2^\d e^{i\phi}- \hat{a}_1^\d \hat{a}_1^\d e^{-i\phi}) \sin\delta \cos\delta).
\end{align}

For all error operators appearing here we have 
$\bra{\bar{i}_N} \hat{a}_1^\dagger\hat{a}_1^\dagger \ket{\bar{j}_N}= 
\bra{\bar{i}_N} \hat{a}_2^\dagger\hat{a}_2^\dagger \ket{\bar{j}_N}= 
\bra{\bar{i}_N} \hat{a}_1^\dagger\hat{a}_2^\dagger \ket{\bar{j}_N}=0$ for all pairs $i,j$. \\

\textbf{Case 5:} $\hat{E}_2^{\dagger}\hat{E}_2$
\begin{align}
\hat{E}_2^{\dagger}\hat{E}_2 =& \kappa_2 t \hat{b}_2^{\dagger} \hat{b}_2+\mathcal{O}((\kappa_i t)^2).
\end{align}

This yields the same condition as obtained from $\hat{E}_1^\dagger\hat{E}_1$.\\

\textbf{Case 6:} $\hat{E}_3^{\dagger}\hat{E}_2$
\begin{align} 
\hat{E}_3^{\dagger}\hat{E}_2 =& (\mathbb{I}- \frac{1}{2} \kappa_2 t \hat{b}^{\dagger}\hat{b})\sqrt{\kappa_1 t}\hat{b}_1^{\dagger} (\mathbb{I}- \frac{1}{2}\kappa_1 t \hat{b}_1\dagger \hat{b}_1)\sqrt{\kappa_2 t} \hat{b}_2 \\
=& \sqrt{\kappa_1 \kappa_2} t \hat{b}_1^\dagger \hat{b}_2 + \mathcal{O}((\kappa_i t)^2)\\
= &\sqrt{\kappa_1 \kappa_2} t (\hat{a}_2^\dagger \hat{a}
_2 e^{i\theta} \sin \delta \cos \delta - \hat{a}_1^\dagger\hat{a}_1 e^{i\theta} \sin \delta \cos \delta + \hat{a}_1^{\dagger}\hat{a}_2 \cos^2 \delta -\hat{a}_2^{\dagger}\hat{a}_1 \sin^2\delta). 
\end{align}
Evaluating the diagonal matrix elements gives
\begin{align}
\bra{\bar{0}_N}\hat{E}_3^{\dagger}\hat{E}_2\ket{\bar{0}_N}&= \sqrt{\kappa_1\kappa_2}t \sum_{m,n}|f_{mn}|^2 (2mN-(2n+1)N)e^{i\theta}\sin\delta\cos\delta,\\
\bra{\bar{1}_N}\hat{E}_3^{\dagger}\hat{E}_2\ket{\bar{1}_N}&= \sqrt{\kappa_1\kappa_2}t \sum_{m,n}|f_{mn}|^2 ((2n+1)N-2mN)e^{i\theta}\sin\delta\cos\delta.
\end{align}
Thus, for the KL condition to be satisfied, if $\delta\neq 0$ we require $\sum_{m,n}|f_{mn}|^2 (2mN-  (2n+1)N)=0$. \\

\textbf{Case 7:} $\hat{E}_3^{\dagger}\hat{E}_3$
\begin{align}
\hat{E}_3^{\dagger}\hat{E}_3 = \kappa_1 t \hat{b}_1^{\dagger} \hat{b}_1 + \mathcal{O}((\kappa_i t)^2).
\end{align}

This produces the same constraint as in the $\hat{E}_1^\dagger\hat{E}_1$ case.

\medskip

\bigskip
\noindent\textbf{Conclusion:} Collecting the above results, the KL conditions for the loss channel to the first order in loss parameter require:
\begin{equation}
    \boxed{\sum_{m,n}|f_{mn}|^2 (m-  n)=\frac{1}{2}}.
    \label{eq:fmn_cond}
\end{equation}

Notably, the binomial instances with $N\ge 2$ and $K=2$ satisfy this condition.

\subsection{Recovery map for purely loss channel}

From Eq.~(\ref{eq:noise_channel_state_evolve}), we see that in the presence of only the loss channel, the state evolution up to first order in the noise strength is given by
 \begin{align}
	\tilde{\mathcal{N}}_L(\bar{\hat{\rho}}) \approx \bar{\hat{\rho}} -\frac{\kappa_1 t}{2}\{\hat{b}_1^\d\hat{b}_1,\bar{\hat{\rho}}\}-\frac{\kappa_2 t}{2}\{\hat{b}_2^\d\hat{b}_2,\bar{\hat{\rho}}\}	
	+\kappa_1 t\;\hat{b}_1\bar{\hat{\rho}}\hat{b}_1^\d+\kappa_2 t\;\hat{b}_2\bar{\hat{\rho}}\hat{b}_2^\d.
\end{align}
For the codewords given in Eq.~(\ref{eq:non-rotated_codewords}), we define the projector onto the code space as 
$\hat{P}_C= \ket{\bar{0}_N}\bra{\bar{0}_N} + \ket{\bar{1}_N}\bra{\bar{1}_N}$.
We now recall the modular measurement POVM defined as
\[
\{\hat{\mathcal{P}}_{00}, \hat{\mathcal{P}}_{01}, \hat{\mathcal{P}}_{10}, (\mathbb{I}-\hat{\mathcal{P}}_{00}-\hat{\mathcal{P}}_{01}-\hat{\mathcal{P}}_{10})\},
\]
where $\hat{\mathcal{P}}_{pq}=\hat{\mathcal{P}}_{p} \otimes \hat{\mathcal{P}}_{q}$ and $\hat{\mathcal{P}}_{p}= \sum_{m}\ketbra{mN+p}$.

Applying this measurement projects the noisy state onto different error syndromes, corresponding to no-jump or photon loss in each mode. We now analyze the resulting state and recovery operation for each syndrome.

\medskip
\noindent $\bullet$ \textbf{ Syndrome $(p,q)=(0,0)$.}
If the measurement outcome corresponds to no-jump $\hat{\mathcal{P}}_{00}$, the projected state becomes
\begin{align}
    \hat{\mathcal{P}}_{00} \tilde{\mathcal{N}}_L(\bar{\hat{\rho}})\hat{\mathcal{P}}_{00}^\d = \bar{\hat{\rho}}- \{ \hat{M}_0^L , \bar{\hat{\rho}} \}
\end{align}
where 
$\hat{M}_0^L= A \hat{a}_1^\d \hat{a}_1 + B \hat{a}_2^\d \hat{a}_2$,
with
$A= \frac{1}{2}(\kappa_1 t \cos^2\delta + \kappa_2 t \sin^2 \delta)$ and
$B= \frac{1}{2}(\kappa_2 t \cos^2\delta + \kappa_1 t \sin^2 \delta)$.

To correct this error we apply the unitary
\begin{align}
    \hat{U}_{00}^L= \exp([\hat{M}_0^L, \hat{P}_C]).
\end{align}
The superscript $L$ corresponds to unitaries for loss noise correction and $D$ in the next section will correspond to unitaries for dephasing noise correction. 
The action of this operation gives 
\begin{align}
    \hat{U}_{00} ^L\hat{\mathcal{P}}_{00}\tilde{\mathcal{N}}_L(\bar{\hat{\rho}})\hat{\mathcal{P}}_{00}^\d \hat{U}_{00}^{L\d}
    &= 
 \big(\mathbb{I}^{\otimes 2} + [\hat{M}_0^L,\hat{P}_C]\big)\big(\bar{\hat{\rho}} - \{\hat{M}_0^L,\bar{\hat{\rho}}\}\big)\big(\mathbb{I}^{\otimes 2} - [\hat{M}_0^L,\hat{P}_C]\big) \nonumber \\
&= \bar{\hat{\rho}} - \{\hat{M}_0^L,\bar{\hat{\rho}}\} + [\hat{M}_0^L,\hat{P}_C]\bar{\hat{\rho}} - \bar{\hat{\rho}}[\hat{M}_0^L,\hat{P}_C] \nonumber \\
    &= \bar{\hat{\rho}} - \{\hat{P}_C \hat{M}_0^L \hat{P}_C, \bar{\hat{\rho}}\} \nonumber \\
    &= \bar{\hat{\rho}} - 2\alpha^2 N (A+B)\bar{\hat{\rho}} ,
    \label{eq:U00}
\end{align}
where we used the equality implied by the KL condition
\[
\alpha^2 N = \sum_{m,n} |f_{mn}|^2 2mN
           = \sum_{m,n} |f_{mn}|^2 (2n+1)N.
\]
We also used that $\hat P_C \bar{\hat{\rho}}= \bar{\hat{\rho}}$ and
$\bra{\bar{i}_N}\hat{M}_0^L\ket{\bar{j}_N} \propto \delta_{ij}$.

\medskip
\noindent $\bullet$\textbf{Syndrome $(p,q)=(0,1)$.}

If the measurement outcome corresponds to $\hat{\mathcal{P}}_{01}$, the projected state becomes
\begin{align}
    \hat{\mathcal{P}}_{01} \tilde{\mathcal{N}}_L(\bar{\hat{\rho}})\hat{\mathcal{P}}_{01}^\d = 2B ~\hat{a}_2 \bar{\hat{\rho}} \hat{a}_2^\d.
\end{align}

We then apply the unitary
\begin{align}
    \hat{U}_{01}^{L}= \ket{\bar{0}_N}\bra{\bar{0}_E}^{(2)}+\ket{\bar{1}_N}\bra{\bar{1}_E}^{(2)},
\end{align}

where
\begin{align}
    \ket{\bar{i}_E}^{(r)}= \frac{\hat{a}_r \ket{\bar{i}_N}}{\sqrt{\alpha^2 N}}.
\end{align}

Here $\left\| \hat{a}_j \ket{\bar{i}_N} \right\| = \sqrt{\alpha^2 N}$ is the norm and $i \in \{0,1\}$ and where $r \in \{1,2\}$. Here we used the KL condition 
$\bra{\bar{i}_N} \hat{a}_r^\d \hat{a}_r \ket{\bar{i}_N}= \alpha^2 N$. Applying the recovery operation yields
\begin{align}
    \hat{U}_{01}^L \hat{\mathcal{P}_{01}}\tilde{\mathcal{N}}_L(\bar{\hat{\rho}})\hat{\mathcal{P}}_{01}^\d \hat{U}_{01}^{L\d
}    &= \big(\ket{\bar{0}_N}\bra{\bar{0}_E}^{(2)}+\ket{\bar{1}_N}\bra{\bar{1}_E}^{(2)} \big)(2B ~ \hat{a}_2 \bar{\hat{
    \rho}} \hat{a}_2^\d)~ \big(\ket{\bar{0}_E}^{(2)} \bra{\bar{0}_N}+\ket{\bar{1}_E}^{(2)}\bra{\bar{1}_N}\big) \nonumber\\
    &= 2B \big(\ket{\bar{0}_N}\bra{\bar{0}_E}^{(2)}+\ket{\bar{1}_N}\bra{\bar{1}_E}^{(2)} \big)~ \big(\sum_{i,j} \alpha^2 N \rho_{ij} \ket{\bar{i}_E}\bra{\bar{j}_E}^{(2)} \big) ~\big(\ket{\bar{0}_E}^{(2)} \bra{\bar{0}_N}+\ket{\bar{1}_E}^{(2)}\bra{\bar{1}_N}\big) \nonumber\\
    &= 2B\alpha^2 N \big(\sum_{i,j} \rho_{i,j} \ket{\bar{i}_N} \bra{\bar{j}_N}\big) \nonumber\\
    &= 2B ~\alpha^2 N \bar{\hat{\rho}},
    \label{eq:U01}
\end{align}

\medskip
\noindent $\bullet$ \textbf{Syndrome $(p,q)=(1,0)$.}

Similarly, if the outcome corresponds to $\hat{\mathcal{P}}_{01}$, we obtain
\begin{align}
    \hat{\mathcal{P}}_{01} \tilde{\mathcal{N}}_L(\bar{\hat{\rho}})\hat{\mathcal{P}}_{01}^\d = 2A ~\hat{a}_1 \bar{\hat{\rho}} \hat{a}_1^\d
\end{align}
and apply the unitary
\begin{align}
    \hat{U}_{10}^L= \ket{\bar{0}_N}\bra{\bar{0}_E}^{(1)}+\ket{\bar{1}_N}\bra{\bar{1}_E}^{(1)} .
\end{align}

Analogously,
\begin{align}
    \hat{U}_{10}^L \hat{\mathcal{P}_{10}}\tilde{\mathcal{N}}_L(\bar{\hat{\rho}})\hat{\mathcal{P}}_{10}^\d \hat{U}_{10}^{L\d}
    &=  2A ~\alpha^2 N.
\label{eq:U10}
\end{align}

If the measurement outcome corresponds to 
$(\mathbb{I}-\hat{\mathcal{P}}_{00}-\hat{\mathcal{P}}_{01}-\hat{\mathcal{P}}_{10})$, we perform no correction since it corresponds to a second-order error.

Thus the recovery channel is
\begin{align}
    \tilde{\mathcal{R}} \sim \{ \hat{U}_{00}^L \hat{\mathcal{P}}_{00}, \hat{U}_{01}^L \hat{\mathcal{P}}_{01}, \hat{U}_{10}^L \hat{\mathcal{P}}_{10}, (\mathbb{I}-\hat{\mathcal{P}}_{00}-\hat{\mathcal{P}}_{01}-\hat{\mathcal{P}}_{10})\}.
\end{align}

Summing Eqs.~(\ref{eq:U00}),(\ref{eq:U01}), and (\ref{eq:U10}) yields
\begin{align}
    \tilde{\mathcal{R}} \circ\tilde{\mathcal{N}}_L(\bar{\hat{\rho}}) = \bar{\hat{\rho}} +  \mathcal{O}\!\left((\kappa_i t)^2\right).
\end{align}

\section{Recovery the two-mode RSB code against dephasing channel up to first order in noise strength}
\label{appsec:recovery_deph}

\subsection{Knill-Laflamme conditions for purely dephasing channel}

We now check the Knill-Laflamme (KL) conditions for the two-mode code under a purely dephasing channel up to first order in $\gamma_i t$.

From Eq.~(\ref{eq:dephasing_kraus_new}), the error operators up to first order are:
\begin{align}
	\hat{\tilde{D}}^{\prime(i)}_0 &= \mathbb{I} - \frac{1}{2}\gamma_i t (\hat{b}^{\dagger}_i\hat{b}_i)^2, \\
	\hat{\tilde{D}}^{\prime(i)}_1 &= \sqrt{\gamma_i t} \hat{b}_i^\dagger\hat{b}_i.
\end{align}

The combined two-mode error operators are:
\begin{align}
    \hat{E}_1 &= \hat{\tilde{D}}^{\prime(1)}_0  \hat{\tilde{D}}^{\prime(2)}_0, \\
    \hat{E}_2 &= \hat{\tilde{D}}^{\prime(1)}_0  \hat{\tilde{D}}^{\prime(2)}_1, \\
    \hat{E}_3 &= \hat{\tilde{D}}^{\prime(1)}_1  \hat{\tilde{D}}^{\prime(2)}_0, \\
    \hat{E}_4 &= \hat{\tilde{D}}^{\prime(1)}_1  \hat{\tilde{D}}^{\prime(2)}_1.
\end{align}

The KL condition requires:
\begin{equation}
    \bra{\bar{i}_N} \hat{E}^{\dagger}_a \hat{E}_b \ket{\bar{j}_N} = C_{ab} \delta_{ij}, \quad a,b=1,\dots,4, \; i,j=0,1.
\end{equation}

We now evaluate each relevant combination up to first order in $\gamma_i t$.

\bigskip
\noindent\textbf{Case 1:} $\hat{E}_1^\dagger \hat{E}_1$
\begin{align}
\hat{E}_1^{\dagger} \hat{E}_1
&= (\mathbb{I}- \frac{1}{2}\gamma_1 t (\hat{b}^{\dagger}_1\hat{b}_1)^2)
(\mathbb{I}- \frac{1}{2}\gamma_2 t (\hat{b}^{\dagger}_2\hat{b}_2)^2)
(\mathbb{I}- \frac{1}{2}\gamma_1 t (\hat{b}^{\dagger}_1\hat{b}_1)^2)
(\mathbb{I}- \frac{1}{2}\gamma_2 t (\hat{b}^{\dagger}_2\hat{b}_2)^2) \\
&= -\gamma_1t (\hat{b}_1^\dagger \hat{b}_1)^2-\gamma_2t (\hat{b}_2^\dagger \hat{b}_2)^2 
+ \mathcal{O}((\gamma_i t)^2) \\
&= -\gamma_1 t \big[(\cos^2\delta~ \hat{a}_1^{\dagger}\hat{a}_1+ \sin^2 \delta \hat{a}_2^{\dagger}\hat{a}_2)^2 + \sin^2 \delta \cos^2 \delta (e^{i\phi}\hat{a}_1^{\dagger}\hat{a}_2 + e^{-i\phi}\hat{a}_2^{\dagger}\hat{a}_1)^2 \big] \nonumber \\
&\quad -\gamma_2 t \big[(\cos^2\delta ~\hat{a}_2^{\dagger}\hat{a}_2+ \sin^2 \delta \hat{a}_1^{\dagger}\hat{a}_1)^2 + \sin^2 \delta \cos^2\delta (-e^{i\phi}\hat{a}_1^{\dagger}\hat{a}_2 + e^{-i\phi}\hat{a}_2^{\dagger}\hat{a}_1)^2 \big] + \text{cross terms}.
\end{align}
For the KL condition, cross terms with unequal numbers of operators in each mode vanish. The diagonal elements give:
\begin{align}
\bra{\bar{0}_N} \hat{E}_1^\dagger \hat{E}_1 \ket{\bar{0}_N} &= -\gamma_1 t \sum_{m,n} |f_{mn}|^2 (2mN \cos^2 \delta + (2n+1)N \sin^2 \delta)^2 \nonumber \\
&\quad -\gamma_2 t \sum_{m,n} |f_{mn}|^2 ((2n+1)N \cos^2\delta + 2mN \sin^2 \delta)^2 \nonumber \\
&\quad - (\gamma_1 +\gamma_2) t \sin^2 \delta \cos^2 \delta \sum_{m,n} |f_{mn}|^2 (2mN((2n+1)N+1)+((2mN+1)(2n+1)N)), \\
\bra{\bar{1}_N} \hat{E}_1^\dagger \hat{E}_1 \ket{\bar{1}_N} &= -\gamma_1 t \sum_{m,n} |f_{mn}|^2 ((2n+1)N \cos^2 \delta + 2mN \sin^2 \delta)^2 \nonumber \\
&\quad -\gamma_2 t \sum_{m,n} |f_{mn}|^2 (2mN \cos^2\delta + (2n+1)N \sin^2 \delta)^2 \nonumber \\
&\quad - (\gamma_1 +\gamma_2) t \sin^2 \delta \cos^2 \delta \sum_{m,n} |f_{mn}|^2 (2mN((2n+1)N+1)+((2mN+1)(2n+1)N)).
\end{align}
Equating these gives the condition:
\begin{equation}
\delta = \pi/4 \quad \text{or} \quad \sum_{m,n}|f_{mn}|^2(2m)^2 = \sum_{m,n}|f_{mn}|^2 (2n+1)^2.
\end{equation}
Off-diagonal terms for $i\neq j$ are reduced to $\bra{\bar{i}_N}\hat{E}_1^{\dagger}\hat{E}_1\ket{\bar{j}_N}$. This is proportional to $ \propto \bra{\bar{i}_N}(\pm e^{i\phi}\hat{a}_1^{\dagger}a_2 +  e^{-i\phi}\hat{a}_2^{\dagger}a_1)^2 \ket{\bar{j}_N}= 0 $ for $N \geq 4$.

\bigskip
\noindent\textbf{Case 2:} $\hat{E}_2^\dagger \hat{E}_1$  
\begin{align}
\hat{E}_2^\dagger \hat{E}_1 &= \sqrt{\gamma_2 t}\hat{b}_2^\dagger \hat{b}_2 (\mathbb{I}- \frac{1}{2}\gamma_1 t (\hat{b}^{\dagger}_1\hat{b}_1)^2)^2 (\mathbb{I}- \frac{1}{2}\gamma_2 t (\hat{b}^{\dagger}_2\hat{b}_2)^2) \\
&= \sqrt{\gamma_2 t}\hat{b}_2^\dagger \hat{b}_2 + \mathcal{O}((\gamma_i t)^2) \\
&= \sqrt{\gamma_2 t} (\cos^2\delta ~\hat{a}_2^{\dagger}\hat{a}_2+ \sin^2 \delta \hat{a}_1^{\dagger}~\hat{a}_1 - \sin \delta \cos\delta (e^{-i\phi}\hat{a}_2^{\dagger}\hat{a}_1 + e^{i\phi}\hat{a}_2^{\dagger}\hat{a}_1)).
\end{align}
Diagonal elements give the same $\delta$ condition as Case 1; off-diagonal terms vanish for any $\delta, \phi$ and $f_{mn}$.

\bigskip
\noindent\textbf{Case 3:} $\hat{E}_3^\dagger \hat{E}_1$  
\begin{align}
\hat{E}_3^\dagger \hat{E}_1 &= \sqrt{\gamma_1 t}\hat{b}_1^\dagger \hat{b}_1 (\mathbb{I}- \frac{1}{2}\gamma_2 t (\hat{b}^{\dagger}_2\hat{b}_2)^2)(\mathbb{I}- \frac{1}{2}\gamma_1 t (\hat{b}^{\dagger}_1\hat{b}_1)^2)(\mathbb{I}- \frac{1}{2}\gamma_2 t (\hat{b}^{\dagger}_2\hat{b}_2)^2) \\
&= \sqrt{\gamma_1 t}\hat{b}_1^\dagger \hat{b}_1 + \mathcal{O}((\gamma_i t)^2),
\end{align}
which is analogous to Case 2.

\bigskip
\noindent\textbf{Case 4:} $\hat{E}_4^\dagger \hat{E}_1$  

\begin{align}
\hat{E}_4^\dagger\hat{E}_1 &=
\sqrt{\gamma_1 \gamma_2}t~\hat{b}_2^\dagger \hat{b}_2 \hat{b}_1^\dagger \hat{b}_1 (\mathbb{I}- \frac{1}{2}\gamma_2 t (\hat{b}^{\dagger}_2\hat{b}_2)^2)(\mathbb{I}- \frac{1}{2}\gamma_1 t (\hat{b}^{\dagger}_1\hat{b}_1)^2)\nonumber\\
&= \sqrt{\gamma_1 \gamma_2}t~\hat{b}_2^\dagger \hat{b}_2 \hat{b}_1^\dagger \hat{b}_1  + 
    \mathcal{O}((\kappa_i t)^2)\nonumber \\
&= \sqrt{\gamma_1 \gamma_2}t (\cos^2\delta ~\hat{a}_2^{\dagger}\hat{a}_2+ \sin^2 \delta \hat{a}_1^{\dagger}~\hat{a}_1+ \sin \delta \cos\delta(-e^{i\phi}\hat{a}_1^{\dagger}a_2 +  e^{-i\phi}\hat{a}_2^{\dagger}a_1)) \times \nonumber\\
&(\cos^2\delta~ \hat{a}_1^{\dagger}\hat{a}_1 +\sin^2~\delta \hat{a}_2^{\dagger}~\hat{a}_2+ \sin \delta \cos\delta(e^{i\phi}\hat{a}_1^{\dagger}a_2+  e^{-i\phi}\hat{a}_2^{\dagger}a_1)).
\end{align}
For the diagonal entries, we get 
\begin{align}
    \bra{\bar{0}_N} \hat{E}_4^\dagger\hat{E}_1 \ket{\bar{0}_N}= \bra{\bar{1}_N} \hat{E}_4^\dagger\hat{E}_1 \ket{\bar{1}_N}= \sqrt{\gamma_1 \gamma_2}t~ (C^2 + 2\alpha^2N)
\end{align}
where $C^2 = \sum_{m,n}|f_{mn}|^2((2mN)^2 + ((2n+1)N)^2)$ and as defined before, $\alpha^2 N = \sum_{m,n} |f_{mn}|^2 2mN= \sum_{m,n} |f_{mn}|^2 (2n+1)N.$
Further we see that the off-diagonals $\bra{\bar{0}_N}\hat{E}_4^\dagger\hat{E}_1 \ket{\bar{1}_N}=0$.

\bigskip
\noindent\textbf{Cases 5--7:} $\hat{E}_2^\dagger \hat{E}_2$, $\hat{E}_3^\dagger \hat{E}_2$, $\hat{E}_3^\dagger \hat{E}_3$  

For these cases we see that, $\hat{E}_2^\dagger \hat{E}_2$,$\hat{E}_3^\dagger \hat{E}_2$ and $\hat{E}_3^\dagger \hat{E}_3$ are similar Case 1, Case 4 and Case 5 respectively. They give the same diagonal conditions as Case 1 and the off-diagonal terms vanish.

\bigskip
\noindent\textbf{Conclusion:}  
For KL conditions to be satisfied up to first order in noise strength, we require that $N \ge 4$ and:
\begin{equation}
\boxed{\delta = \pi/4} \quad \text{or} \quad \boxed{\sum_{m,n}|f_{mn}|^2 (m^2 - n^2 - n) = \frac{1}{4}}.
\end{equation}
We note that the KL conditions for second order in noise strength $\kappa_i t$ are not satisfied. 

\subsection{Recovery map for purely dephasing channel}

From Eq.~\ref{eq:noise_channel_state_evolve}, we see that in the presence of a purely dephasing channel, the state evolution up to first order in the noise strength is given by
\begin{align}
	\tilde{\mathcal{N}}_D(\bar{\hat{\rho}}) \approx \bar{\hat{\rho}} -\frac{\gamma_1 t}{2}\{(\hat{b}_1^\d\hat{b}_1)^2,\bar{\hat{\rho}}\}-\frac{\gamma_2 t}{2}\{(\hat{b}_2^\d\hat{b}_2)^2,\bar{\hat{\rho}}\}
	+\gamma_1 t\;(\hat{b}_1^\d\hat{b}_1)\bar{\hat{\rho}}(\hat{b}_1^\d\hat{b}_1)+\gamma_2 t\;(\hat{b}_2^\d\hat{b}_2)\bar{\hat{\rho}}(\hat{b}_2^\d\hat{b}_2).
\end{align}

We employ the following modular measurement,
\[
\{ \hat{\mathcal{P}}_{00}, \hat{\mathcal{P}}_{-1,1}, \hat{\mathcal{P}}_{1,-1}, (\mathbb{I}-\hat{\mathcal{P}}_{00}-\hat{\mathcal{P}}_{1,-1}-\hat{\mathcal{P}}_{-1,1}) \}.
\]

Applying this measurement projects the noisy state onto different error syndromes corresponding to either no jump or photon exchange between the two modes.  
For the no-jump component, we will see that an additional projective measurement is required for full recovery.

Since the modular measurements are performed in the non-transformed mode basis—chosen for experimental feasibility in microwave cavities—we still require the following condition to hold, even for a purely dephasing channel:
\[
\sum_{m,n} |f_{mn}|^2 2mN
= \sum_{m,n} |f_{mn}|^2 (2n+1)N .
\]
This condition ensures that the recovery map described below remains valid.

We note, however, that for a purely dephasing channel the KL conditions are always satisfied when $\delta=\pi/4$. Therefore, there should exist an appropriate basis in which the recovery map can be constructed without imposing this additional constraint. Nevertheless, since the above condition coincides with the one obtained from the KL analysis of the loss channel, it is convenient to employ the same family of codes for both noise processes. With this choice, the recovery map presented below applies directly to the dephasing case as well.

\medskip
\noindent $\bullet$ \textbf{Syndrome $(p,q)=(0,0)$.}

If the measurement outcome corresponds to the no-jump event $\hat{\mathcal{P}}_{00}$, the projected state becomes

\begin{align}
    \hat{\mathcal{P}}_{00} \tilde{\mathcal{N}}_D(\bar{\hat{\rho}})\hat{\mathcal{P}}_{00}^\d
    =
    \bar{\hat{\rho}}- \{ \hat{M}_0^D , \bar{\hat{\rho}} \}+ \hat{M}_1^{D} \hat{\rho}\hat{M}_1^{D\d}.
\end{align}

Here
\[
\hat{M}_0^D= \frac{(\gamma_1 +\gamma_2)t}{8}\big[(\hat{a}_1^\d\hat{a}_1+\hat{a}_2^\d\hat{a}_2)^2 + \hat{a}_2^\d \hat{a}_1 \hat{a}_1^\d \hat{a}_2 +\hat{a}_1^\d \hat{a}_2 \hat{a}_2^\d \hat{a}_1 \big],
\]
and
\[
\hat{M}_1^{D}= \frac{\sqrt{(\gamma_1+ \gamma_2)t}}{2}(\hat{a}_1^\d \hat{a}_1 + \hat{a}_2^\d \hat{a}_2).
\]

To correct the first term, we apply the unitary
\begin{align}
    \hat{U}_{00}^D= \exp([\hat{M}_0^D, \hat{P}_C]).
\end{align}

The action of this operation yields, up to $O(\gamma_i t)$,
\begin{align}
    \hat{U}_{00}^D \hat{\mathcal{P}}_{00}\tilde{\mathcal{N}}_D(\bar{\hat{\rho}})\hat{\mathcal{P}}_{00}^\d \hat{U}_{00}^{D\d
}    &= 
 \big(\mathbb{I}^{\otimes 2} + [\hat{M}_0^D,\hat{P}_C]\big)
 \big(\bar{\hat{\rho}} - \{\hat{M}_0^D,\bar{\hat{\rho}}\} +\hat{M}_1^{D}\bar{\hat{\rho}}\hat{M}_1^{D\d}\big)
 \big(\mathbb{I}^{\otimes 2} - [\hat{M}_0^D,\hat{P}_C]\big) \nonumber \\
    &= \bar{\hat{\rho}} - \{\hat{P}_C \hat{M}_0^D \hat{P}_C , \bar{\hat{\rho}}\}+ \hat{M}_1^{D}\bar{\hat{\rho}}\hat{M}_1^{D\d} \nonumber \\
    &= \bar{\hat{\rho}} - \frac{(\gamma_1+\gamma_2)t}{4}(4\beta_0^2 + C^2 +\alpha^2N)  \bar{\hat{\rho}}+\hat{M}_1^{D}\bar{\hat{\rho}}\hat{M}_1^{D\dagger}.
\end{align}

In deriving this expression we used
\[
\bra{\bar{0}_N}\hat{M}_0^D\ket{\bar{0}_N}
=
\bra{\bar{1}_N}\hat{M}_0^D\ket{\bar{1}_N}
=
C^2+ 4\beta_0^2 + 2\alpha^2N,
\]
where we defined
\[
\beta^2 = \sum_{m,n} |f_{mn}|^2~ 2m(2n+1)N^2 .
\]
To correct the remaining term $\hat{M}_1^{D}\bar{\hat{\rho}}\hat{M}_1^{D\d}$, we perform an additional projective measurement
\[
\hat{P}_{CE}=\{\hat{P}_C +
\hat{P}_1^{'}, \hat{P}_E +\hat{P}_2^{'}\},
\]
where
\begin{align}
    \hat{P}_E =\ket{\bar{0}}_E^{\perp}\bra{\bar{0}}_E^{\perp}+\ket{\bar{1}}_E^{\perp}\bra{\bar{1}}_E^{\perp}.
\end{align}

Here we have defined the orthogonal states
\[
\ket{\bar{i}}_E^{\perp}= (\hat{a}_1^\d \hat{a}_1 + \hat{a}_2^\d \hat{a}_2)\ket{\bar{i}_N},
\]
which can be written as
\begin{align}
    \ket{\bar{i}}_E^{\perp}
    =
    \frac{1}{\sqrt{C_2 - C_1^2}}
    \big(C_1 \ket{\bar{i}_N} -\ket{\bar{i}}_E \big).
\end{align}
where $C_1=\bra{\bar{0}_N}(\hat{a}_1^\d \hat{a}_1 + \hat{a}_2^\d \hat{a}_2)\ket{\bar{0}_N}= \bra{\bar{1}_N}(\hat{a}_1^\d \hat{a}_1 + \hat{a}_2^\d \hat{a}_2)\ket{\bar{1}_N}= 2\alpha^2N$ and $C_2 = \bra{\bar{0}_N}(\hat{a}_1^\d \hat{a}_1 + \hat{a}_2^\d \hat{a}_2)^2\ket{\bar{0}_N}= \bra{\bar{1}_N}(\hat{a}_1^\d \hat{a}_1 + \hat{a}_2^\d \hat{a}_2)^2\ket{\bar{1}_N}= \sum_{m,n}|f_{mn}|^2\big((2mN)^2 + ((2n+1)N)^2 + 4m(2n+1)N^2= C^2+ 2\beta_0^2$. This ensures the states are correctly normalised. 
The projectors $\hat{P}_i^{'}$ above are complementary projectors to ensure that the POVMs sum to identity. \\

If the measurement outcome corresponds to $\hat{P}_C$ we do nothing, while for the $\hat{P}_E$ outcome we apply
\[
\hat{U}_E^{D}= \ket{\bar{0}_N}\bra{\bar{0}_E}^{\perp}+ \ket{\bar{1}_N}\bra{\bar{1}_E}^{\perp}.
\]
The action of these operators on the term $\hat{M}_1^{D}\bar{\hat{\rho}}\hat{M}_1^{D\d}$ is
\begin{align}
    \hat{P}_{C} \hat{M}_1^{D} \bar{\hat{\rho}}\hat{M}_1^{D\d}\hat{P}_{C}+ \hat{U}_E^{D}\hat{P}_{E} \hat{M}_1^{D} \bar{\hat{\rho}}\hat{M}_1^{D\d}\hat{P}_{E}\hat{U}_E^{D\d}
    &=\frac{(\gamma_1+\gamma_2)t}{4}(C^2 + 2\beta_0^2)\bar{\hat{\rho}} .
\end{align}

Combining the above results, the contribution from the no-jump sector becomes
\begin{align} \hat{P}_C \hat{U}_{00}^D \hat{\mathcal{P}}_{00}\tilde{\mathcal{N}}_D(\bar{\hat{\rho}})\hat{\mathcal{P}}_{00}^\d \hat{U}_{00}^{D\d} \hat{P}_C &+ \hat{U}_E^{D}\hat{P}_E \hat{U}_{00}^D \hat{\mathcal{P}}_{00}\tilde{\mathcal{N}}_D(\bar{\hat{\rho}})\hat{\mathcal{P}}_{00}^\d \hat{U}_{00}^{D\d}\hat{P}_E \hat{U}_E^{D\d}\nonumber \\ &= \bar{\hat{\rho}} +\frac{(\gamma_1+\gamma_2)t}{4}\big(-4\beta_0^2 -C^2 -2\alpha^2 N + C^2 +2\beta_0^2) \nonumber\\ &=\bar{\hat{\rho}} +\frac{(\gamma_1+\gamma_2)t}{4}\big(-2\beta_0^2 -2\alpha^2 N\big). 
\label{eq:U00D}
\end{align}
\medskip
\noindent $\bullet$ \textbf{Syndrome $(p,q)=(1,-1)$.}

If the measurement outcome corresponds to a photon exchange with gain in the first mode and loss in the second mode, the projected state becomes
\begin{align}
    \hat{\mathcal{P}}_{1,-1} \tilde{\mathcal{N}}_D(\bar{\hat{\rho}})\hat{\mathcal{P}}_{1,-1}^\d
    =
    \frac{(\gamma_1 +\gamma_2)t}{4}(\hat{a}_1^\d \hat{a}_2 \bar{\hat{\rho}} \hat{a}_2^\d\hat{a}_1).
\end{align}

To correct this error we apply the unitary
\begin{align}
    \hat{U}_{1,-1}= \ket{\bar{0}_N}\bra{\bar{0}}_{1,-1}  +\ket{\bar{1}_N}\bra{\bar{1}}_{1,-1}.
\end{align}

Here
\[
\ket{\bar{i}}_{1,-1} =
\frac{1}{\sqrt{\beta}}(\hat{a}_1^\d \hat{a}_2)\ket{\bar{i}_N},
\]
with
\[
\beta=
\beta_0^2 +\alpha^2 N.
\]

The action of the recovery operation then gives
\begin{align}
    \hat{U}_{1,-1}\hat{\mathcal{P}}_{1,-1} \tilde{\mathcal{N}}_D(\bar{\hat{\rho}})\hat{\mathcal{P}}_{1,-1}^\d \hat{U}_{1,-1}^\dagger
    =
    \frac{(\gamma_1 +\gamma_2)t}{4}(\beta_0^2+ \alpha^2N)\bar{\hat{\rho}}.
    \label{eq:U1-1}
\end{align}

\medskip
\noindent $\bullet$ \textbf{Syndrome $(p,q)=(-1,1)$.}

For this outcome the projected state is
\begin{align}
    \hat{\mathcal{P}}_{-1,1} \tilde{\mathcal{N}}_D(\bar{\hat{\rho}})\hat{\mathcal{P}}_{-1,1}^\d
    =
    \frac{(\gamma_1 +\gamma_2)t}{4}(\hat{a}_2^\d \hat{a}_1 \bar{\hat{\rho}} \hat{a}_1^\d\hat{a}_2).
\end{align}

We apply the analogous unitary
\begin{align}
    \hat{U}_{-1,1}= \ket{\bar{0}_N}\bra{\bar{0}}_{-1,1}  +\ket{\bar{1}_N}\bra{\bar{1}}_{-1,1},
\end{align}
which yields
\begin{align}
    \hat{U}_{-1,1}\hat{\mathcal{P}}_{-1,1} \tilde{\mathcal{N}}_D(\bar{\hat{\rho}})\hat{\mathcal{P}}_{-1,1}^\d \hat{U}_{-1,1}^\dagger
    =
    \frac{(\gamma_1 +\gamma_2)t}{4}(\beta_0^2+ \alpha^2N)\bar{\hat{\rho}}.
    \label{eq:U-11}
\end{align}

The recovery operation can be written as
\begin{align}
	\tilde{\mathcal{R}}_D \sim \Big\{ & \hat{P}_C\hat{U}_{00}\hat{P}_{00}, \; \hat{U}_E^{D}\hat{P}_E\hat{U}_{00}\hat{P}_{00}, \nn \\
	& \hat{U}_{1,-1}\hat{P}_{1,-1}, \; \hat{U}_{-1,1}\hat{P}_{-1,1} \Big\}
	\cup
	\{ \hat{P}_{lm} \,|\, \text{all other }(l,m) \}.
\end{align}

Summing Eqs.~(\ref{eq:U00D}), (\ref{eq:U1-1}), and (\ref{eq:U-11}), its action on the noisy state is therefore
\begin{align}\tilde{\mathcal{R}}_D\circ\tilde{\mathcal{N}}_D(\bar{\hat{\rho}})& = \bar{\hat{\rho}} + \frac{(\gamma_1 +\gamma_2)t}{4}(-2\beta_0^2 -2\alpha^2 N) + 2\frac{(\gamma_1 +\gamma_2)t}{4}(\beta_0^2 +\alpha^2 N) + \mathcal{O}((\gamma_i t)^2) \nonumber \\ &=\bar{\hat{\rho}} +\mathcal{O}((\gamma_i t)^2). \end{align}

\section{Near-optimal fidelity under loss and dephasing channels}
\label{app:near-optimal}
In the original work by H.~Michael \emph{et al.}~\cite{qec3}, the performance of single-mode binomial codes in the case $N=K$ was investigated for various values of $N$. For a fixed noise strength $\kappa t$, they showed that there exists an optimal binomial code with a finite value of $N$ that minimizes the entanglement fidelity among different binomial codes. Motivated by this result, we investigate how this behavior extends to our two-mode binomial codes.

In the two-mode case, the dimension of the codespace grows rapidly with increasing $N$ and $K$, making it computationally prohibitive to obtain the optimal recovery map via semidefinite programming. To address this challenge, we employ the near-optimal channel fidelity introduced in Ref.~\cite{nearopt}, which provides an optimization-free metric applicable to arbitrary noise channels and quantum codes. This quantity yields a two-sided bound on the optimal code performance and depends only on the Knill--Laflamme matrix.

The Knill--Laflamme quantum error-correction matrix is defined as
\begin{equation}
M_{[\mu l][\nu k]} = \bra{\mu_L}\hat{N}_l^\dagger \hat{N}_k \ket{\nu_L},
\end{equation}
where $\ket{\nu_L}$ are the logical codewords and $\{\hat{N}_k\}$ are the Kraus operators of the noise channel $\mathcal{N}$. The near-optimal fidelity is given by
\begin{equation}
\tilde{F}^{\mathrm{opt}} = \frac{1}{d_L^2}
\left\| \Tr_L \sqrt{M} \right\|_F^2,
\end{equation}
where $(\Tr_L B)_{l,k} = \sum_{\mu} B_{[\mu l],[\mu k]}$, $d_L$ is the logical dimension, and $\|\cdot\|_F$ denotes the Frobenius norm. This quantity provides the following two-sided bound on the optimal fidelity $F^{\mathrm{opt}}$:
\begin{equation}
\frac{1}{2}\bigl(1-\tilde{F}^{\mathrm{opt}}\bigr)
\leq 1 - F^{\mathrm{opt}}
\leq 1-\tilde{F}^{\mathrm{opt}}.
\end{equation}
When a code performs well against a given noise channel, the near-optimal fidelity serves as an accurate approximation to the true optimal fidelity. We therefore use these near-optimal bounds to assess the performance of our dual-rail binomial codes (DRBC) for larger values of $N$ and $K$. The performance under photon loss is shown in Fig.~\ref{fig:near-optimal}(a). For the loss channel, we find that the behavior of the two-mode binomial codes closely mirrors that of the single-mode case. In particular, for small noise strengths, the performance improves with increasing $N$, while for larger noise strengths there exists an optimal value of $N$ that minimizes the infidelity. This crossover behavior in the single-mode case was previously analyzed in Ref.~\cite{qec3}.

However for purely dephasing channel, as we showed in the previous section that the performance improves for $N\ge 4$ as the KL conditions are satisfied up to first order in noise strength, we see a significant improvement unlike the case of the single-mode codes. In the next section~\ref{sec:phase-meas}, we will further show that the KL conditions for continuous representation of dephasing Kraus operators become closer to being satisfied as $N \rightarrow \infty$. The numerical results for this analytic proof is given here in the Fig.~\ref{fig:near-optimal}. For DBRC with $K=2$ for optimal encoding angles, we vary $N$ and we plot the optimal fidelity obtained from SDP calculations. We further validate it with results from near-optimal recovery. In this plot, $\mathcal{F}_E$ is the entanglement fidelity which is related to the average gate fidelity as $\mathcal{F}=(2\mathcal{F}_E+1)/3$ for the qubit encoding. 

\begin{figure}[H]
	\centering
       \includegraphics[width=\textwidth]{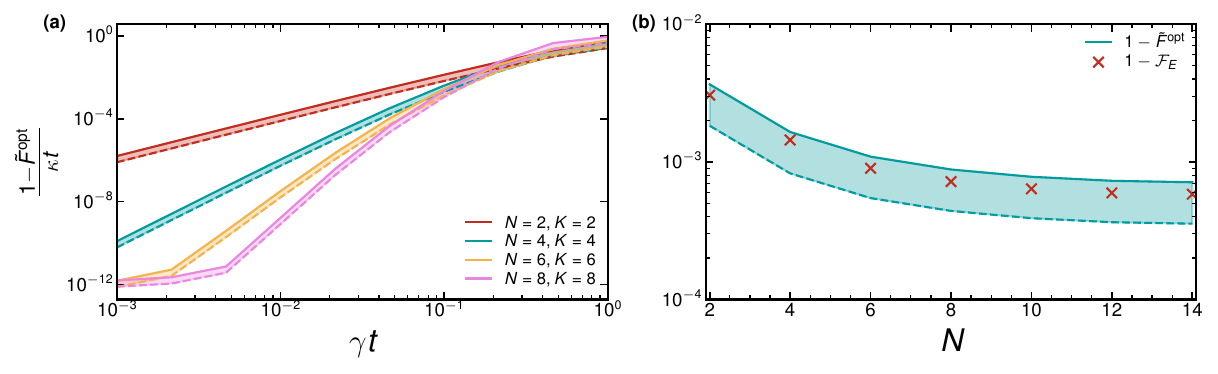}
       \caption{The solid lines in the figure represent the upper bound of the near-optimal infidelity given by $1-\tilde{F}^{\text{opt}}$ and the dashed line represents the lower bound given by  $\frac{1}{2}(1- \tilde{F}^{\text{opt}})$. The subfigure (a) is evaluated for DRBC $N=K$ for only loss as a function of noise strength $\kappa_1= \kappa_2= \kappa$. In (b), by keeping $K=2$ and $\gamma_1 t=\gamma_2 t= \gamma t= 0.01$ fixed, we vary $N$ to study the performance with dephasing noise. The infidelity decreases with increase in $N$; in contrast to the single-mode case.}
	\label{fig:near-optimal}
\end{figure}

\section{More understanding on performance under dephasing for two-mode RSB using phase measurements and KL conditions of dual codewords}
\label{sec:phase-meas}
In Sec.~\ref{appsec:recovery_deph}, we analyzed the effect of the Gaussian dephasing channel using its countable Kraus representation (Eq.~(\ref{eq:dephasing_kraus_new})) and constructed a recovery map valid up to first order in the noise strength. We showed that an exact first-order recovery map for the two-mode code can be obtained when the encoding mode is fixed to $\delta = \pi/4$ or $3\pi/4$ (independent of $\phi$), provided that the coefficients $f_{mn}$ satisfy the condition given in Eq.~(\ref{eq:fmn_cond}).

In this section, we instead consider the Gaussian dephasing channel acting on the two modes in terms of its continuous Kraus representation~\cite{Grimsmo_2020}. Using this representation, we first examine under what conditions the Knill–Laflamme (KL) criteria are satisfied. We then analyze the distinguishability of the codewords under phase measurements in order to gain further insight into the performance of these codes in the presence of dephasing noise.

 \subsection{Knill-Laflamme matrix for dual codewords under dephasing noise}\label{appsec:overlap_KL}

The effective noisy channel is given by $\tilde{\mathcal{N}} = \hat{U}_{BS}^\d\mathcal{N}\hat{U}_{BS}$ where the noise channel $\mathcal{N}$   can be expanded in terms of the elements of the continuous Kraus operators, 
\begin{align}
	\mathcal{N}\sim\{\exp(i\hat{A}(\theta_1,\theta_2))= e^{i\theta_1\hat{a}_1^\d\hat{a}_1+ i\theta_2\hat{a}_2^\d\hat{a}_2}\;|\;\theta_1,\theta_2\in[0,2\pi)\}.
\end{align}
For the effective channel $\tilde{\mathcal{N}}$, the transformed Kraus operators take the form:
\begin{align}\label{eq:A-def}
	\tilde{\hat{A}}(\theta_1,\theta_2) &= \hat{U}_{BS}^\d (\theta_1\hat{a}_1^\d\hat{a}_1+ \theta_2\hat{a}_2^\d\hat{a}_2) \hat{U}_{BS}  \nn\\
    &=\hat{a}_1^\d\hat{a}_1 (\theta_1 \cos^2\delta +\theta_2\sin^2\delta) + \hat{a}_2^\d\hat{a}_2 (\theta_2 \cos^2\delta +\theta_1\sin^2\delta)\nn\\ &+ (\theta_1-\theta_2)(e^{-i\phi}\hat{a}_1^\d\hat{a}_2+e^{i\phi}\hat{a}_2^\d\hat{a}_1)\sin\delta\cos\delta.
\end{align}

It is worth noting that the countable Kraus operators for the dephasing channel, $\{(\hat{a}^\d\hat{a})^l|l\in\mathbb{N}\}$~\cite{Albert_2019,Grimsmo_2020}, can be expanded in terms of the continuous Kraus operators shown above; the latter are nothing but the rotation operators in the phase space, which allows us to have an intuitive understanding of the \textit{distortion} of information in bosonic systems under dephasing channels. In fact, the dephasing noise can be thought of as diffusion between the probability distributions of the phase measurement outcomes corresponding to the dual codewords.
If we choose the encoding mode to be given by $\delta=\pi/4$, we obtain 
 \begin{align}
  	\tilde{\hat{A}}(\theta_1,\theta_2) &= \frac{1}{2}\l(\hat{a}_1^\d\hat{a}_1+\hat{a}_2^\d\hat{a}_2\r) (\theta_1 +\theta_2) +  \frac{1}{2}(\theta_1-\theta_2)(e^{-i\phi}\hat{a}_1^\d\hat{a}_2+e^{i\phi}\hat{a}_2^\d\hat{a}_1)\\
  	 & = \frac{1}{2}\l(\hat{n}_1+\hat{n}_2\r) (\theta_1 +\theta_2) +  \frac{1}{2}(\theta_1-\theta_2)(\hat{G}^+_{12} \cos\phi - \hat{G}^-_{12}\sin\phi).
  \end{align}
Since $[(\hat{n}_1+\hat{n}_2),\hat{G}^{\pm}_{12}] = 0$, the noise operator $\exp(i\hat{A}(\theta_1,\theta_2))$ decomposes as
\begin{align}
\label{eq:decomp}
	\exp(\frac{i}{2}(\theta_1-\theta_2)(\hat{G}^+_{12} \cos\phi - \hat{G}^-_{12}\sin\phi))\exp(\frac{i}{2}\l(\hat{n}_1+\hat{n}_2\r) (\theta_1 +\theta_2)).
\end{align}
The off-diagonal entries of the Knill-Laflamme matrix in the dual bases for the two error operators with angles $(\theta'_1,\theta'_2)$ and $(\theta''_1,\theta''_2)$ can be now simplified using Eq.~(\ref{eq:decomp}) as
\begin{align}
\label{eq:KL-continuous-operators-dephasing}
	\bra{-_N}\exp(-i\tilde{\hat{A}}(\theta'_1,\theta'_2)) \exp(i\tilde{\hat{A}}(\theta''_1,\theta''_2)) \ket{+_N}
    =\bra{-_N}\exp(i\tilde{\hat{A}}(\theta_1,\theta_2))\ket{+_N}
\end{align}
where $\theta_i = \theta''_i-\theta'_i$. Simplifying this further, we obtain
\begin{align}
		\bra{-_N}&\exp(\frac{i}{2}(\theta_1-\theta_2)(\hat{G}^+_{12} \cos\phi - \hat{G}^-_{12}\sin\phi))\exp(\frac{i}{2}\l(\hat{n}_1+\hat{n}_2\r) (\theta_1 +\theta_2))\ket{+_N}.\nn\\
		%=&\bra{-_N}\exp(\frac{i}{2}(\theta_1-\theta_2)(\hat{G}^+_{12} \cos\phi - \hat{G}^-_{12}\sin\phi))\exp(\frac{i}{2}\l(\hat{n}_1+\hat{n}_2\r) (\theta_1 +\theta_2))\ket{+_N}. \nn
\end{align}
We further see that when $\phi=0$, %e
\begin{align}
   \bra{-_N}&\exp(\frac{i}{2}(\theta_1-\theta_2)\hat{G}^+_{12})\exp(\frac{i}{2}\l(\hat{n}_1+\hat{n}_2\r) (\theta_1 +\theta_2))\ket{+_N} \nn \\
   =& \bra{-_N} e^{-i\frac{\pi}{2}\hat{G}^-_{12}}e^{i\frac{\pi}{2}\hat{G}^-_{12}} \exp(\frac{i}{2}(\theta_1-\theta_2)\hat{G}^+_{12})\exp(\frac{i}{2}\l(\hat{n}_1+\hat{n}_2\r) (\theta_1 +\theta_2)) e^{-i\frac{\pi}{2}\hat{G}^-_{12}}e^{i\frac{\pi}{2}\hat{G}^-_{12}}\ket{+_N} \nn \\
    =&-\bra{-_N}e^{i\frac{\pi}{2}\hat{G}^-_{12}}\exp(\frac{i}{2}(\theta_1-\theta_2)\hat{G}^+_{12})e^{-i\frac{\pi}{2}\hat{G}^-_{12}}\exp(\frac{i}{2}\l(\hat{n}_1+\hat{n}_2\r) (\theta_1 +\theta_2))\ket{+_N}\nn\\
    =&-\bra{-_N}e^{-i\pi\hat{n}_2}e^{i\pi\hat{n}_2}\exp(-\frac{i}{2}(\theta_1-\theta_2)\hat{G}^+_{12}) \exp(\frac{i}{2}\l(\hat{n}_1+\hat{n}_2\r) (\theta_1 +\theta_2))e^{-i\pi\hat{n}_2}e^{i\pi\hat{n}_2}\ket{+_N}\nn\\
		=&-\bra{-_N}e^{-i\pi\hat{n}_2}e^{i\pi\hat{n}_2}\exp(-\frac{i}{2}(\theta_1-\theta_2)\hat{G}^+_{12}) \exp(\frac{i}{2}\l(\hat{n}_1+\hat{n}_2\r) (\theta_1 +\theta_2))e^{-i\pi\hat{n}_2}e^{i\pi\hat{n}_2}\ket{+_N}\nn\\
		=&-\bra{-_N}\exp(\frac{i}{2}(\theta_1-\theta_2)\hat{G}^+_{12})\exp(\frac{i}{2}\l(\hat{n}_1+\hat{n}_2\r) (\theta_1 +\theta_2))\ket{+_N}.
        \label{eq:deph_off_diag_1}
\end{align}
Here we have used the fact that $e^{i\frac{\pi}{2}\hat{G}^-_{12}}$ acts as the logical $X$ operator on the codespace and $N$ is an even number. We have also used the commutator relation $[\hat{G}^{-}_{12}, \hat{G}^{+}_{12}]= 2i(\hat{a}_1^\d \hat{a}_1 -\hat{a}_2^\d \hat{a}_2)$ from which we also obtain the relation $e^{i\frac{\pi}{2}\hat{G}^-_{12}} \hat{G^{+}_{12}}e^{-i\frac{\pi}{2}\hat{G}^-_{12}}= -\hat{G^+_{12}}$. 
Thus from the above equality in Eq.(\ref{eq:deph_off_diag_1}), we can conclude
\begin{align}
     \bra{-_N}\exp(\frac{i}{2}(\theta_1-\theta_2)\hat{G}^+_{12})\exp(\frac{i}{2}\l(\hat{n}_1+\hat{n}_2\r) (\theta_1 +\theta_2))\ket{+_N}=0.\label{eq:dephasing_offdiagonal}
\end{align}
For the existence of exact recovery, we further require that the diagonal entries are equal. However for this channel, we see that they differ as shown below:
\begin{align}
	&\bra{+_N}\exp(\frac{i}{2}(\theta_1-\theta_2)(\hat{G}^+_{12} \cos\phi - \hat{G}^-_{12}\sin\phi))\exp(\frac{i}{2}\l(\hat{n}_1+\hat{n}_2\r) (\theta_1 +\theta_2))\ket{+_N}\nn\\
	=&\bra{-_N}e^{-i\frac{\pi}{N}\hat{n}_2}\exp(\frac{i}{2}(\theta_1-\theta_2)(\hat{G}^+_{12} \cos\phi - \hat{G}^-_{12}\sin\phi))\exp(\frac{i}{2}\l(\hat{n}_1+\hat{n}_2\r) (\theta_1 +\theta_2))e^{i\frac{\pi}{N}\hat{n}_2}\ket{-_N}\nn\\
	=&\bra{-_N}\exp(\frac{i}{2}(\theta_1-\theta_2)\l(\hat{G}^+_{12} \cos(\frac{\pi}{N}-\phi) + \hat{G}^-_{12}\sin(\frac{\pi}{N}-\phi)\r))\exp(\frac{i}{2}\l(\hat{n}_1+\hat{n}_2\r) (\theta_1 +\theta_2))\ket{-_N},
    \label{eq:dephasing_diagonal}
\end{align}  
where we have used the relations $e^{-i\frac{\pi}{N}\hat{n}_2} ~\hat{G}^{\pm}_{12}~ e^{i\frac{\pi}{N}\hat{n}_2}= \hat{G}^{\pm}_{12} \cos\tfrac{\pi}{N} ~\pm~\hat{G}^{\mp}_{12}\sin\tfrac{\pi}{N}$ for the proof. 
However, from the above expression in Eq.(\ref{eq:dephasing_diagonal}), we see that the two diagonal entries can be brought arbitrarily close to each other by setting $\phi=(2m+1)\frac{\pi}{2N}$ for $m \in \mathbb{Z}$, and taking the limit $N\to\infty$. We also present numerical results in Fig.~\ref{fig:near-optimal}(b) for DRBC to support these findings: as $N$ increases for a fixed $K$, the average gate infidelity decreases, indicating improved performance. Moreover, we see that for all $\theta_1=\theta_2$ corresponding to the case of correlated dephasing, the entries given in Eq.(\ref{eq:dephasing_diagonal}) are exactly equal. This justifies the existence of the recovery circuit in Fig. 3 in the main text. 
Note that the above discussion holds for any of the general two-mode instantiations of our codes, namely given by Eqs.(4) and (5).

We exemplify this analytical understanding with  numerical calculation of the entanglement infidelity for the case of the DRBC with $N=2$ and $K=2$. As shown in Fig.\ref{fig:enter-label}, the infidelity is minimized for the  optimal encoding angles, $\delta=\pi/4,3\pi/4$ and $\phi=\pi/4$. We have also numerically checked that for $N=4, K=2$, the infidelity is minimized for the  optimal encoding angles, $\delta=\pi/4,3\pi/4$ and $\phi=\pi/8$ (not shown here). 

\begin{figure}[H]
    \centering
    \includegraphics[width=\linewidth]{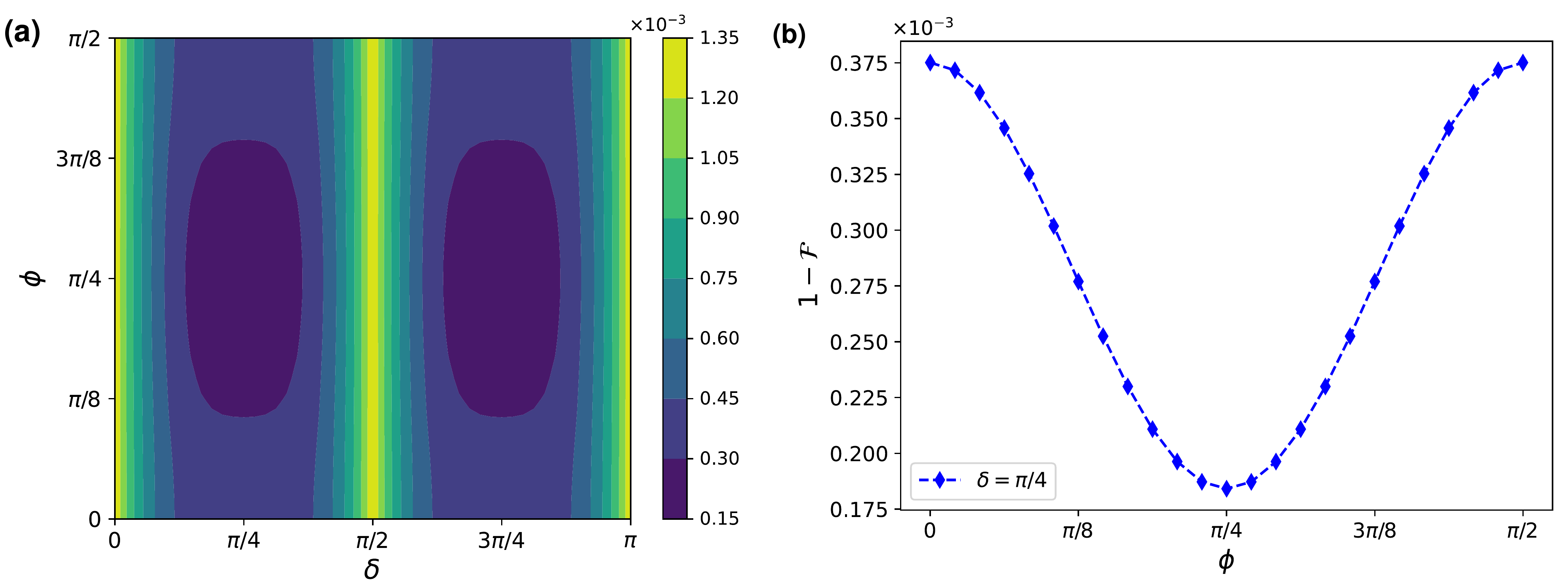}
    \caption{ (a) Entanglement infidelity as a function of $\delta$ and $\phi$, showing minima for $\delta=\pi/4,3\pi/4$ for $N=2, K=2$ DRBC against dephasing $\gamma_1 t= \gamma_2 t= 10^{-3}$. (b) The variation of entanglement infidelity as a function of $\phi$ is very small (of the order $10^{-3}$). We however still observe that the lowest value is obtained from $\phi= \pi/4$ which is $\pi/(2N)$ for $N=2$ and $m=0$.}
    \label{fig:enter-label}
\end{figure}

\subsection{Distinguishing the dual codewords and understanding the phase trajectories}
\label{sec:dual-words}
In this Section, we will see how the two-mode code is distinct in performance under dephasing as compared to a single mode code in terms of the phase distinguishability of the codewords. 
 Similar to the case of single-mode RSB, the dual codewords, $\ket{\pm_N}$ encoded in the two-mode RSB, are related by a rotation in either mode, up to a global phase: $\exp(i\frac{\pi}{N}\hat{a}_2^\d\hat{a}_2)\ket{\pm_N} = \ket{\mp_N};\;\exp(i\frac{\pi}{N}\hat{a}_1^\d\hat{a}_1)\ket{\pm_N}= -\ket{\mp_N}$.
  
  The \textit{joint phase measurement} in each of the modes formally can be written as a set of \textit{positive operator-valued measures} (POVM): $\{\hat{\Pi}(\phi_1)\otimes\hat{\Pi}(\phi_2)\;|\;\phi_1,\phi_2\in[0,2\pi)\}$, where $\hat{\Pi}(\phi)$ is defined e.g. in Ref.\cite{udupa2025performance}. The outcomes of such measurement is $(\phi_1,\phi_2)$. In the absence of any noise, outcomes $(0,\pi/N)\mod (2\pi/N)$ or $(\pi/N,0)\mod (2\pi/N)$ indicate that the logical state is $\ket{-_N}$, while the measurement outcomes $(0,0)\mod (2\pi/N)$ or $(\pi/N,\pi/N)\mod (2\pi/N)$  indicate that the logical state is $\ket{+_N}$ (Fig.~\ref{fig:phase-dual}). 

  \begin{figure}[H]
  \centering
         \subfigure[]{\includegraphics[width=0.45\textwidth]{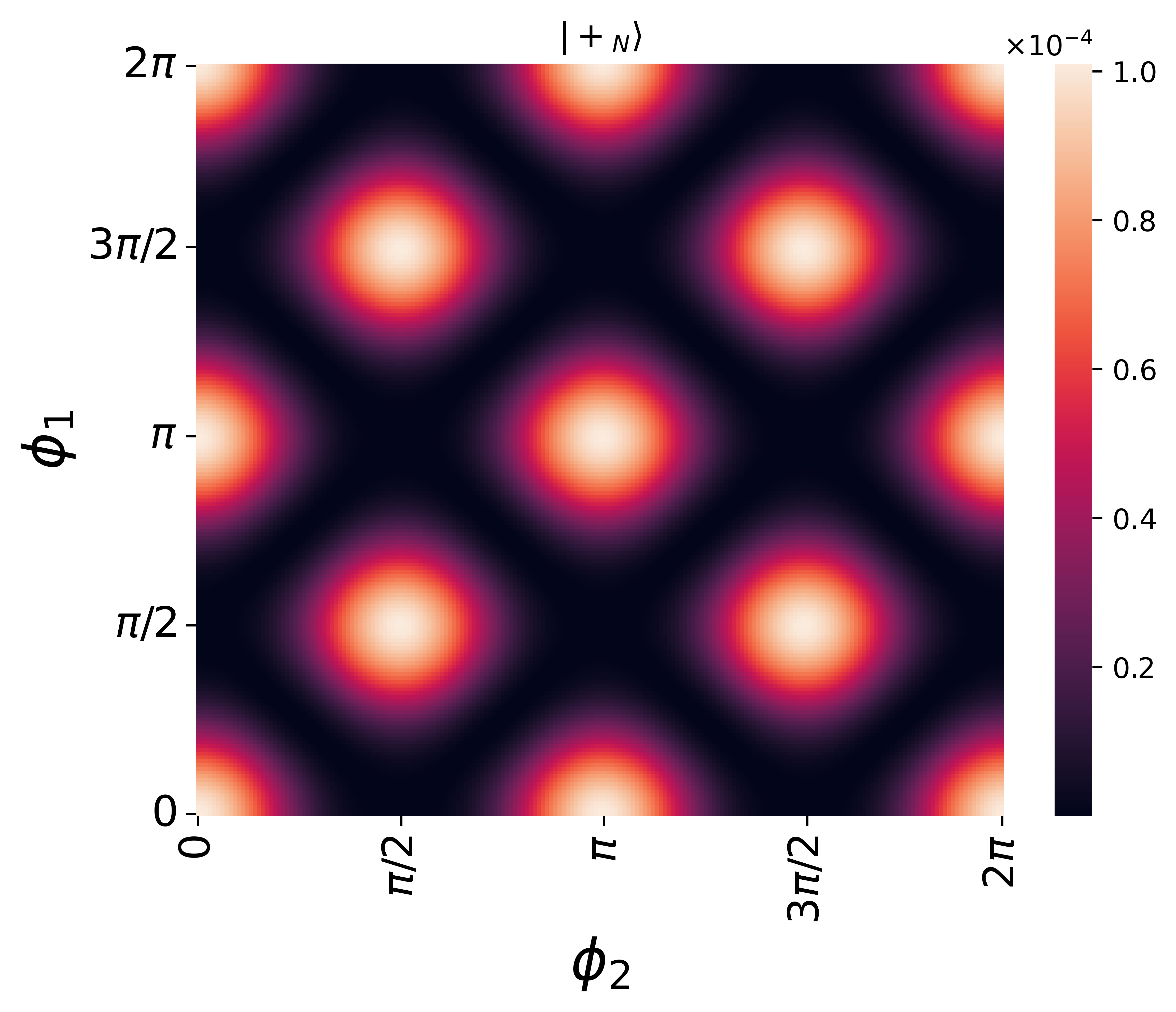}}
        \subfigure[]{\includegraphics[width=0.42\textwidth]{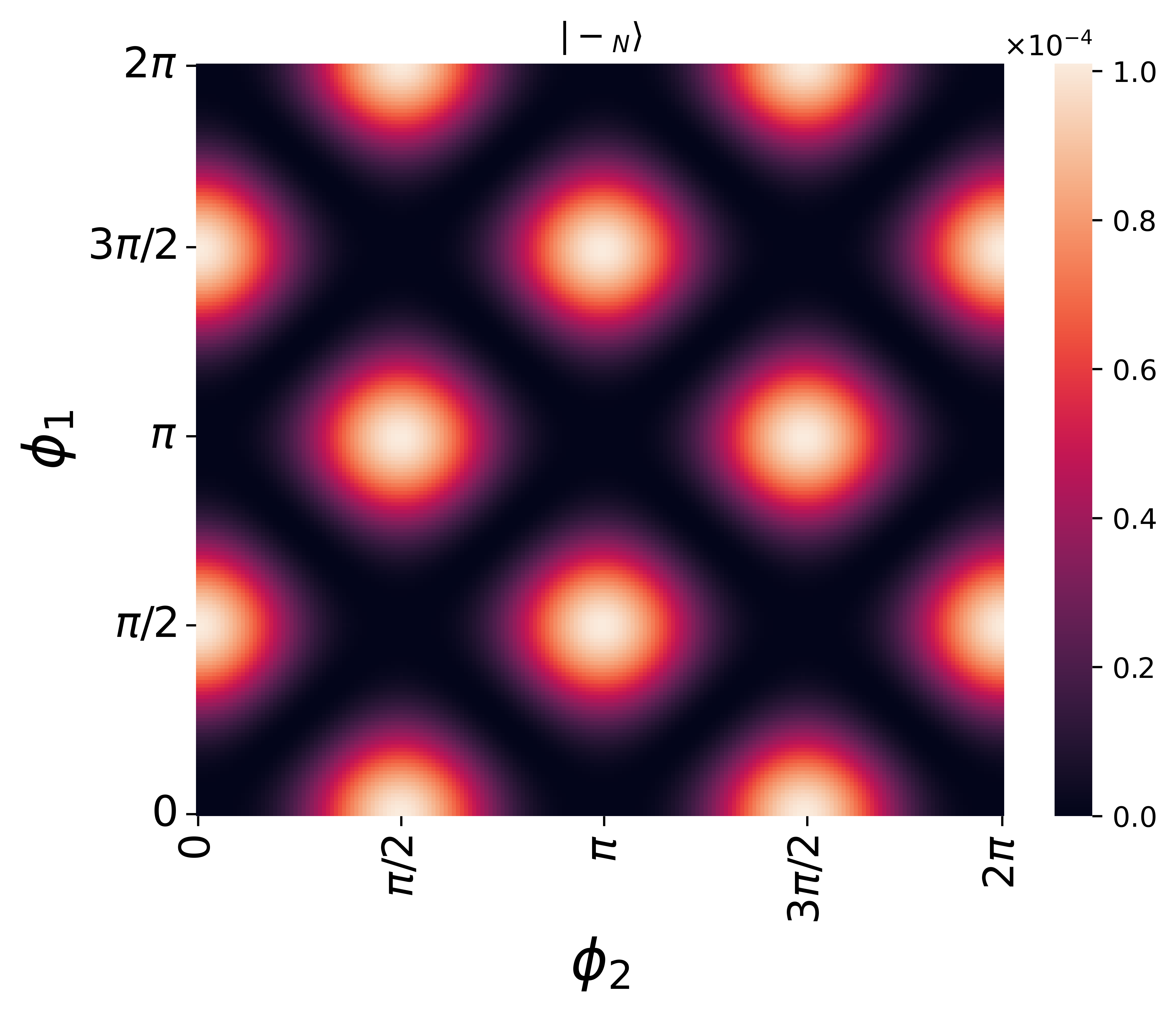}}
  	  	\caption{Probability distribution of phase measurement outcome for the dual basis encoded in $K=2,N=2$ two-mode code.}
  	\label{fig:phase-dual}
  \end{figure}
  %Insert heat-map figure

As we have previously found out, for $\delta = 0$ in Eq.~(\ref{eq:A-def}), the physical error operators, $\exp(i\hat{A}(0,\pi/N))$ and $\exp(i\hat{A}(\pi/N,0))$ acts as logical $Z$ operator. Hence, the probability distribution of the outcomes of the joint phase measurement can be used to distinguish the codewords for \textit{small enough} rotation errors only, parametrised by $\theta_1,\theta_2 \ll \pi/N$. This allows us to distinguish the dual codewords with high accuracy for small enough errors.  However for large rotation errors, for example $(\pi/N,0), (0,\pi/N)$, the joint probability distribution for the dual codewords maps into one another, as these errors act as logical $Z$ operator.
%
\begin{comment}
\begin{figure}[H]
  	\centering
         \subfigure[]{\includegraphics[width=0.45\textwidth]{plus0.png}}
        \subfigure[]{\includegraphics[width=0.42\textwidth]{minus0.png}}
  
  	\caption{\textit{Probability distribution of phase measurement outcome for the dual basis encoded in $N=2,K=2$ two-mode code under a small rotation $\exp(i\hat{A}(0.1\pi,0))$for the encoding axis given by $\delta = 0$}}
  	\label{fig:noised=0}
  \end{figure}
  %Insert heat-map figures for noisy codewords.
\end{comment}

 On the other hand, if we consider the case, $\delta=\pi/4$, from Eq.~(\ref{eq:A-def}) we have
  \begin{align}
  	\tilde{\hat{A}}(\theta_1,\theta_2) &= \frac{1}{2}\l(\hat{a}_1^\d\hat{a}_1+\hat{a}_2^\d\hat{a}_2\r) (\theta_1 +\theta_2) +  \frac{1}{2}(\theta_1-\theta_2)(e^{-i\phi}\hat{a}_1^\d\hat{a}_2+e^{i\phi}\hat{a}_2^\d\hat{a}_1)\\
  	 & = \frac{1}{2}\l(\hat{n}_1+\hat{n}_2\r) (\theta_1 +\theta_2) +  \frac{1}{2}(\theta_1-\theta_2)(\hat{G}_+ \cos\phi - \hat{G}_-\sin\phi).
  \end{align}
As we know that $[(\hat{n}_1+\hat{n}_2),\hat{G}_\pm] = 0$, the noise operators decomposes as
\begin{align}
\tilde{\hat{A}}(\theta_1,\theta_2) =\exp(\frac{i}{2}(\theta_1-\theta_2)(\hat{G}_+ \cos\phi - \hat{G}_-\sin\phi))\exp(\frac{i}{2}\l(\hat{n}_1+\hat{n}_2\r) (\theta_1 +\theta_2)).
\end{align}
As shown in Sec.~\ref{sec:phase-meas}, and in particular in Eq.(\ref{eq:dephasing_offdiagonal}), for $\delta = \pi/4$ and $\phi=0$, the overlap between the codewords as it evolves under the dephasing Kraus operators, remains zero:
\begin{align}\label{eq:zero-overlap}
		\bra{-_N}\exp(i\tilde{\hat{A}}(\theta_1,\theta_2))\ket{+_N} = 0 \quad \forall \theta_1,\theta_2.
\end{align}
This is in stark contrast with the single-mode and $\delta=0$ two-mode cases, where the overlap can be non-zero.
Moreover, if we look at the joint probability distribution of phase measurement outcomes of the dual codewords as shown in Fig.~\ref{fig:noised=pi/4}, we find that their support are largely mutually exclusive for all rotation errors. This allows us to distinguish the dual codewords with high accuracy, even for large rotation errors. 
However we re-emphasise that exact error-correction is generally still not possible since as shown in the previous section, the diagonal entries of the Knill-Laflamme equations are not equal (Eq.(\ref{eq:dephasing_diagonal})).

\begin{figure}[H]
	\centering
         \subfigure[]{\includegraphics[width=0.45\textwidth]{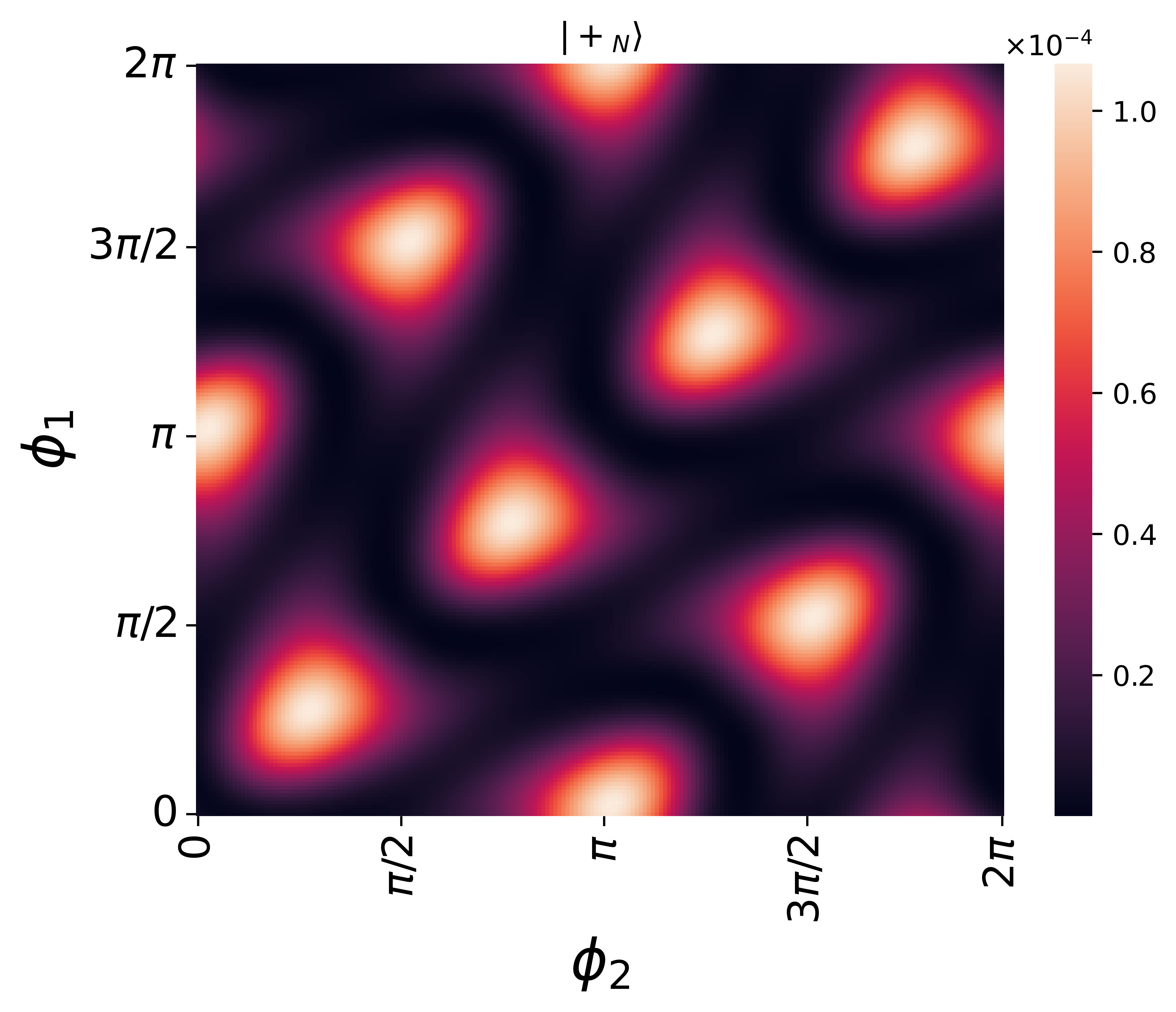}}
        \subfigure[]{\includegraphics[width=0.42\textwidth]{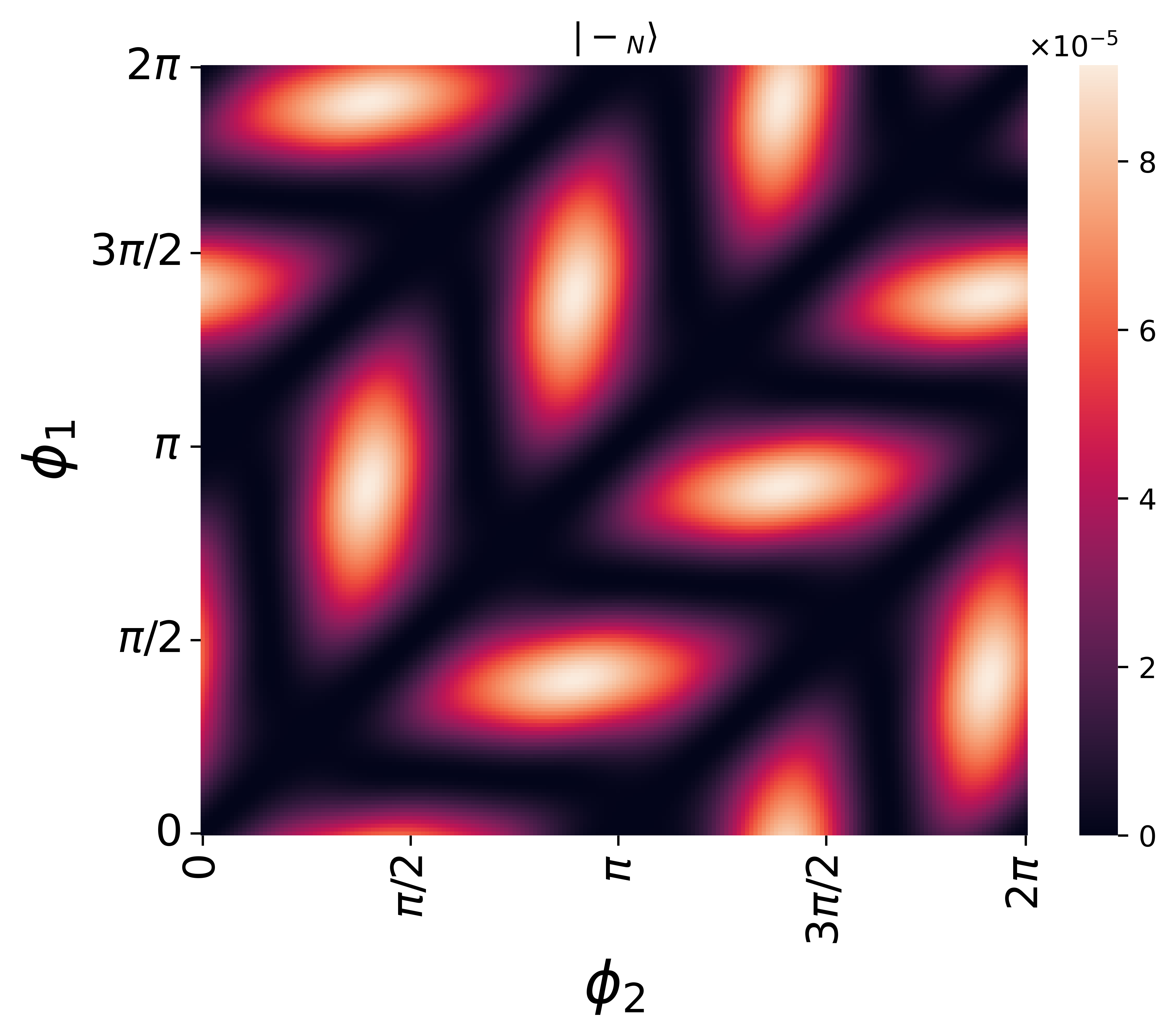}}
	
	\caption{Probability distribution of phase measurement outcome for the dual basis encoded in $K=2,N=2$ two-mode code under a rotation $\exp(i\hat{A}(0.2\pi,0))$ for the encoding axis given by $\delta = \frac{\pi}{4},\phi = 0$}.
	\label{fig:noised=pi/4}
\end{figure}
%Insert heat-map figures for noisy codewords. 

It should also be noted that Eq.~(\ref{eq:zero-overlap}) holds true for Gaussian dephasing alone, and does not imply that the evolution of trace distance between the codewords under an arbitrary dephasing channel is always unity. In general, the noisy codewords are not perfectly distinguishable for an arbitrary dephasing channel. 
A future work may explore whether such non-overlapping distribution of the outcomes of phase measurements can be used to design an error-correcting circuit similar to the teleportation circuit due to Knill~\cite{Grimsmo_2020,timo}. Finally, also note that for DRBC, as the energy of the binomial codes (parameterised by $K$ for binomial codes) increases, the probability distribution of the outcomes for the dual codewords has vanishing overlap \cite{Grimsmo_2020}, suggesting a better performance against the dephasing noise can be found with increasing $K$ (while with losses a sweet-spot in average energy is to be expected). 

\section{Performance of two-mode rotation symmetric bosonic codes under the influence of random telegraph noise}\label{appsec:rtn}

Here we present numerical results on the performance of  simple instances of our DRBC ($K=2,N=2$), for different choices of the encoding angles, against Random Telegraph Noise. $\xi$ is the switching rate and $\nu$ is the amplitude of the RTN. We refer to Ref.\cite{udupa2025performance} for a thorough definition of the noise, along with the parameters.  The plot in Fig. \ref{fig:N=2,K=2,rtn,two} shows that the optimal encoding angles yield an increased performance against RTN noise, and that such performance beats the case of the corresponding single-mode binomial code.
\begin{figure}[H]
	\centering
	\includegraphics[width=14cm]{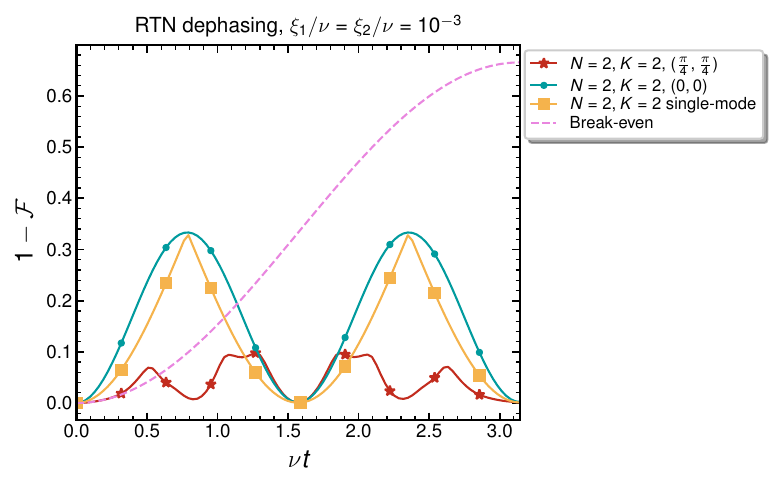}
	\caption{A comparison between the optimal performance of the single-mode binomial code, $K=2,N=2$ and the simplest instance of the two-mode order-$N$ RSB code, $K=2,N=2$ inner binomial concatenated with outer dual rail. Specifically we investigate the performance of the code, upon a beam-splitting action with angle $(\pi/4,\pi/4)$ after the preparation of the logical state in the modes $\hat{a}_1,\hat{a}_2$.}
	\label{fig:N=2,K=2,rtn,two}
\end{figure}

\section{Exact recovery circuit for correlated dephasing}\label{appsec:exact-corr}

Here we show that correlated dephasing can be corrected perfectly, and we derive the corresponding recovery map. Since the beam-splitting operation $\hat{U}_{BS}$ commutes with the Kraus operators of correlated dephasing, i.e., $[(\hat{n}_1+\hat{n}_2),U_{BS}(\delta,\phi)] = 0$ for each $\delta,\phi$, the analysis holds independently of the choice of encoding angles. 

First, we wish to study the action of the controlled operation 
\[
CX_{NL} = \exp\!\left(i\frac{\pi}{2N}\hat{n}_1\otimes\hat{G}^{-}_{34}\right)
\]
on the four-mode Fock states appearing in the tensor product of two codewords provided in the main text, which we report here for convenience:
\begin{align}
\label{eq:codewords-general-two-modes-0}
    \ket{0_N} &= \hat{U}_{BS}\sum_{m,n}f_{mn}\ket{2mN,(2n+1)N}\\
    \ket{1_N} &= \hat{U}_{BS}\sum_{m,n}f_{mn}\ket{(2n+1)N,2mN}.
   \label{eq:codewords-general-two-modes-1}
\end{align}
%in Eqs.~(\ref{eq:codewords-general-two-modes-0}) and (\ref{eq:codewords-general-two-modes-1}).

To do so, we first look explicitly at the action of beam-splitter operations of the type in Eq.(\ref{eq:beam-splitter-explicit}) on states of the form $|rN,sN \rangle$, with $r$ and $s$ two integer numbers.
The operator $e^{i \delta \hat G^-_{12}}$ corresponds to setting the angle $\phi = 2 \pi k$ with $k$ integer in Eq.(\ref{eq:beam-splitter-explicit}). 
Its action can be expressed as 
\begin{align}
e^{i \delta \hat G^-_{12}} |rN, sN\rangle &= \left( e^{i \delta \hat G^-_{12}} \frac{(\hat a_1^\dagger)^{rN}} {\sqrt{(rN)!}} e^{-i \delta \hat G^-_{12}} \right) \left( e^{i \delta \hat G^-_{12}}   \frac{(\hat a_2^\dagger)^{sN}}{\sqrt{(sN)!}} e^{-i \delta \hat G^-_{12}} \right) |0, 0\rangle \nonumber \\
&=  \frac{(\hat{a}_1^\dagger \cos\delta +\hat{a}_2^\dagger \sin\delta )^{rN}} {\sqrt{(rN)!}}  \frac{(\hat{a}_2^\dagger \cos\delta -\hat{a}_1^\dagger \sin\delta)^{sN}}{\sqrt{(sN)!}}  |0, 0\rangle,
\end{align}
where  we have used Eq.(\ref{eq:beam-splitter-explicit2}) for $\phi = 2 \pi k$, and the fact  that the operator $e^{i \delta \hat G^-_{12}} $ acts as the identity on the two-mode vacuum.

The previous equation reduces for some values of $\delta$ to the two relevant cases:
\begin{align}
\text{for } \delta &= \pi m \rightarrow e^{i \delta \hat G^-_{12}} |rN, sN\rangle= (-1)^{mN(r+s)} \frac{(\hat a_1^\dagger)^{rN}} {\sqrt{(rN)!}} \frac{(\hat a_2^\dagger)^{sN}}{\sqrt{(sN)!}}  |0, 0\rangle =  |rN, sN\rangle \nonumber \\
\text{for } \delta &= (2n+1) \frac{\pi}{2} \rightarrow e^{i \delta \hat G^-_{12}} |rN, sN\rangle = (-1)^{n N(r-s)} \frac{(\hat a_1^\dagger)^{rN}} {\sqrt{(rN)!}} \frac{(\hat a_2^\dagger)^{sN}}{\sqrt{(sN)!}}  |0, 0\rangle =  |rN, sN\rangle,
\end{align}
where we have used that $N$ is even.
As a consequence of this, the action of the controlled operation on the four-mode Fock states appearing in the tensor product of two codewords of the form in Eqs.~(\ref{eq:codewords-general-two-modes-0}) and (\ref{eq:codewords-general-two-modes-1}) gives
\begin{align}
\label{eq:terms-in-codewords-general-two-modes}
 CX_{NL} \ket{2mN,(2n+1)N} \otimes \ket{2pL,(2q+1)L} &=  \ket{2mN,(2n+1)N} \otimes e^{i \pi m \hat G^-_{34}}\ket{2pL,(2q+1)L} \nonumber \\
&= \ket{2mN,(2n+1)N} \ket{2pL,(2q+1)L}\\
 CX_{NL} \ket{2mN,(2n+1)N} \otimes \ket{(2q+1)L, 2pL} &=  \ket{2mN,(2n+1)N} \otimes e^{i \pi m \hat G^-_{34}}\ket{(2q+1)L, 2pL} \nonumber\\
&= \ket{2mN,(2n+1)N} \ket{(2q+1)L, 2pL}\\
  CX_{NL} \ket{(2n+1)N,2mN} \otimes  \ket{2pL,(2q+1)L} &=\ket{(2n+1)N,2mN} \otimes e^{i \frac{\pi}{2}(2n +1) \hat G^-_{34}}  \ket{2pL,(2q+1)L}\nonumber\\
  &=\ket{(2n+1)N,2mN} \otimes\ket{(2q+1)L,2pL}\\
   CX_{NL} \ket{(2n+1)N,2mN} \otimes \ket{(2q+1)L, 2pL} &=\ket{(2n+1)N,2mN} \otimes e^{i \frac{\pi}{2}(2n +1) \hat G^-_{34}} \ket{(2q+1)L, 2pL}\nonumber\\
  &=\ket{(2n+1)N,2mN} \otimes\ket{2pL,(2q+1)L}.
\end{align}

%\begin{align}
%\label{eq:beam-splitter-explicit2}
%	\hat{b}_1 &= \hat{U}^\d_{BS}\hat{a}_1\hat{U}_{BS} = \hat{a}_1\cos\delta +\hat{a}_2e^{-i\phi}\sin\delta\\
%	\hat{b}_2 &= \hat{U}^\d_{BS}\hat{a}_2\hat{U}_{BS} = \hat{a}_2\cos\delta -\hat{a}_1e^{i\phi}\sin\delta.
%\end{align}

Summing the terms in Eq.(\ref{eq:terms-in-codewords-general-two-modes}) over the indices $m,n,p,$ and $q$ in the above expressions, one finds that $CX_{NL}\ket{0}_N\otimes\ket{a}_L = \ket{0}_N\otimes\ket{a}_L$ and $CX_{NL}\ket{1}_N\otimes\ket{a}_L = \ket{1}_N\otimes\ket{a \oplus a}_L$, indicating that it acts as a controlled-$X$ gate on the codespace of the two-mode codes. 

Using these expressions, the map corresponding to the circuit in Fig. 3, with $CX_{LN} = \exp\!\left(i\frac{\pi}{2N}\hat{n}_3\otimes\hat{G}^{-}_{21}\right)$, is given by

\begin{align}
	&\mathcal{R}(\mathcal{N}(\hat{\rho})\otimes\ketbra{0}_L)\nn\\
	= &CX_{LN}CX_{NL} \left(\mathcal{N}(\hat{\rho})\otimes \ketbra{0}_L\right) CX^\dagger_{NL}CX^\dagger_{LN}\nn\\
	=&  \sum_{a,b=0,1}\rho_{ab} CX_{LN}CX_{NL} \left(\int_{-\infty}^\infty \dd\phi\;p_t(\phi)e^{i\phi(\hat{n}_1+\hat{n}_2)}\;\ketbra{a}{b}_N\;e^{-i\phi(\hat{n}_1+\hat{n}_2)}\otimes \ketbra{0}_L\right) CX^\dagger_{NL}CX^\dagger_{LN}\nn\\
	=&\sum_{a,b=0,1}\rho_{ab}  \left(\int_{-\infty}^\infty \dd\phi\;p_t(\phi)e^{i\phi(\hat{n}_1+\hat{n}_2)}\;CX_{LN}\ketbra{a}{b}_N\otimes \ketbra{a}{b}_L CX^\dagger_{LN} \;e^{-i\phi(\hat{n}_1+\hat{n}_2)}\right). 
    \label{eq:corr_mid}
\end{align}

Here we have used the fact that the $CX$ gate commutes with the correlated dephasing operator, since $[(\hat{n}_i+\hat{n}_j),\hat{G}^{-}_{ij}] = 0$, and that the controlled-$X$ operation satisfies $CX_{NL}\ket{a}_N\ket{0}_L= \ket{a}_N\ket{a}_L$.  

For a given $a$ and $b$, we further find that
\begin{align}
	&CX_{LN}\ket{a}_N\otimes \ket{a}_L ~~\bra{b}_N \otimes\bra{b}_L CX_{LN}^{\dagger} \nn\\
	=& \ket{(a \oplus a)}_N \otimes \ket{a}_L  ~~ \bra{(b \oplus b)}_N \otimes \bra{b}_L\nn\\
	=& \ketbra{0}{0}_N \otimes \ketbra{a}{b}_L.
\end{align}
%\begin{align}
%	&CX_{LN}\ket{a}_L\otimes \ket{a}_N ~~\bra{b}_N \otimes\bra{b}_L CX_{LN}^{\dagger} \nn\\
%	=& \ket{a}_L \otimes \ket{(a \oplus a)}_N~~ \bra{(b \oplus b)}_N \otimes \bra{b}_L\nn\\
%	=& \ketbra{0}{0}_N \otimes \ketbra{a}{b}_L.
%\end{align}

Substituting this into Eq.~(\ref{eq:corr_mid}), we obtain 
\begin{align}
	\mathcal{R}(\mathcal{N}(\hat{\rho})\otimes\ketbra{0}_L) & =  \left(\int_{-\infty}^\infty \dd\phi\;p_t(\vec{\phi})e^{i\phi(\hat{n}_1+\hat{n}_2)}\;\ketbra{0}{0}_N\;e^{-i\phi(\hat{n}_1+\hat{n}_2)}\right)\otimes\sum_{a,b = 0,1}\rho_{ab} \ketbra{a}{b}_L \nn \\
   = & \mathcal{N}(\ketbra{0}{0}_N)\otimes\sum_{a,b = 0,1}\rho_{ab} \ketbra{a}{b}_L.
\end{align}
This holds for an arbitrary choice of the two-mode codewords in Eqs.~(4) and (5), not only for the case of binomial codes.

\section{Hadamard gate and $T$ gate}\label{appsec:hgate-tgate}

In this section, we include the circuits that implement the Hadamard and $T$ gates via gate teleportation. The schemes are direct extensions of those proposed for single-mode RSB codes~\cite{Grimsmo_2020} to the two-mode setting (see Fig.~\ref{fig:hgate-tgate}). We first show how the controlled rotation ($CROT$) gate acts as a logical controlled-$Z$ ($CZ$) gate on our encoded two-mode codespace.  

The $CROT$ gate is defined as
\[
CROT_{NL} = \exp\!\left(\frac{i\pi}{NL}\hat{n}_1\otimes \hat{n}_3\right),
\]
where $\hat{n}_1$ and $\hat{n}_3$ are the photon number operators associated with the first modes of the control and target rails, respectively. The gate acts on the four-mode Fock states appearing in the tensor product of two codewords as 

\begin{align}
CROT_{NL} \ket{2mN,(2n+1)N} \otimes \ket{2pL,(2q+1)L} &=  e^{i \pi 2m 2p} \ket{2mN,(2n+1)N} \otimes \ket{2pL,(2q+1)L} \nonumber \\
&= \ket{2mN,(2n+1)N} \ket{2pL,(2q+1)L}\\
CROT_{NL}  \ket{2mN,(2n+1)N} \otimes \ket{(2q+1)L, 2pL} &= e^{i \pi 2m (2q+1)} \ket{2mN,(2n+1)N} \otimes \ket{(2q+1)L, 2pL} \nonumber\\
&= \ket{2mN,(2n+1)N} \ket{(2q+1)L, 2pL}\\
CROT_{NL} \ket{(2n+1)N,2mN} \otimes  \ket{2pL,(2q+1)L} &=  e^{i \pi(2n+1)2p}\ket{(2n+1)N,2mN} \otimes \ket{2pL,(2q+1)L}\nonumber\\
  &=\ket{(2n+1)N,2mN} \otimes\ket{2pL,(2q+1)L}\\
CROT_{NL} \ket{(2n+1)N,2mN} \otimes \ket{(2q+1)L, 2pL} &=e^{i \pi(2n+1)(2q+1)}\ket{(2n+1)N,2mN} \otimes \ket{(2q+1)L, 2pL}\nonumber\\
  &= -\ket{(2n+1)N,2mN} \otimes\ket{(2q+1)L, 2pL}.
\end{align}

Summing over the indices, one can verify that $CROT_{NL}$ acts as a controlled-$Z$ gate on the logical codespace $\{\ket{a_N} \otimes \ket{b_L},~ a,b \in \{0,1\}\}$.

To illustrate the Hadamard-gate protocol in Fig.~\ref{fig:hgate-tgate} (a), we first prepare an auxiliary two-mode encoded qubit in the state $\ket{+_M} = \frac{1}{\sqrt{2}}(\ket{0_M}+\ket{1_M})$ and entangle it with the state $\ket{\psi_N} = \alpha \ket{0_N}+ \beta \ket{1_N}$ to be teleported:
\begin{equation}
    CROT \ket{\psi_N} \otimes \ket{+_M} = \alpha \ket{0_N}\ket{+_M} + \beta\ket{1_N}\ket{-_M}.
\end{equation}

Rewriting the order-$N$ state in the $\pm$ basis and the order-$M$ state in the computational basis, we obtain
\begin{equation}
    CROT \ket{\psi_N} \otimes \ket{+_M} = \frac{\alpha}{2}(\ket{+_N}+ \ket{-_N})\otimes (\ket{0_M}+ \ket{1_M}) +  \frac{\beta}{2}(\ket{+_N}- \ket{-_N})\otimes (\ket{0_M}-\ket{1_M}).
\end{equation}

Measuring the first (control) rail in the $\pm$ basis yields the output state on the target rail as $\alpha \ket{+_M} + \beta \ket{-_M}$ for the outcome ``$+$'' and $\alpha \ket{+_M} - \beta \ket{-_M}$ for the outcome ``$-$''. Combining both cases, with $i=0$ for ``$+$'' and $i=1$ for ``$-$'', the output state is
\begin{equation}
\bar{X}^i \bar{H} \ket{\psi_M},
\end{equation}
where $\bar{H}$ acts as the logical Hadamard gate on the two-mode codespace.

The same reasoning applies for the $T$-gate circuit in Fig.~\ref{fig:hgate-tgate} (b). Writing $\ket{T_M} = \frac{1}{\sqrt{2}}(\ket{0_M}+e^{\frac{i\pi}{4}}\ket{1_M})$, one obtains the logical $\bar{T}$ or $\bar{T}\bar{X}$ gate depending on whether the measurement outcome is ``$+$'' or ``$-$'', respectively. 

\begin{figure*}[h!!]
         \centering
         \subfigure[]{\includegraphics[width=0.4\textwidth]{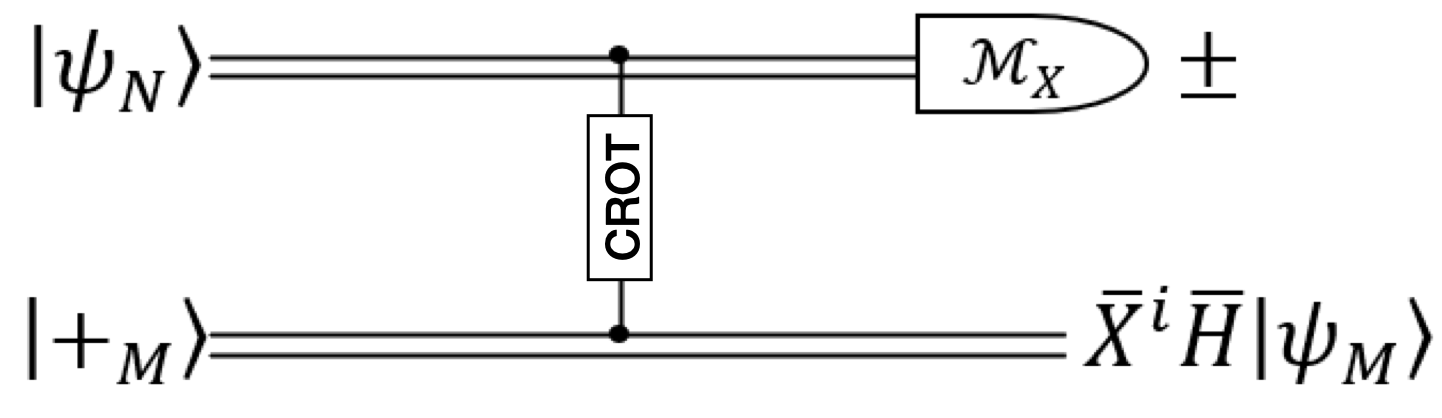}}
        \subfigure[]{\includegraphics[width=0.4\textwidth]{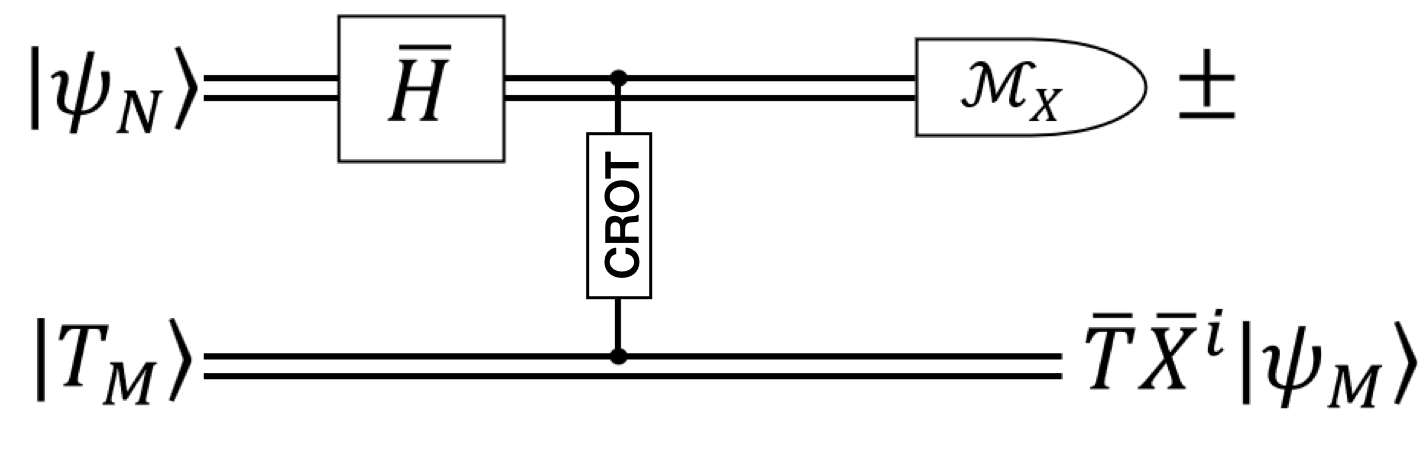}}
         \caption{Implementation of the (a) Hadamard gate and (b) $T$-gate for the two-mode rotation-symmetric codes using gate teleportation.}
         \label{fig:hgate-tgate}
\end{figure*}

\end{widetext}

\end{document}